\documentclass{aa}  

\usepackage{lipsum}
\usepackage{graphicx}
\usepackage{txfonts}
\usepackage{booktabs}
\usepackage{xcolor} 
\usepackage{comment}
\usepackage{multirow}
\usepackage{tikz}

\newcommand{\vect}[1]{\boldsymbol{#1}}

\newcommand{\mpch}[0]{$h^{-1}\,\mathrm{Mpc}$}

\newcommand{\hmpc}[0]{$h\,\mathrm{Mpc}^{-1}$}

\newcommand{\wonj}[0]{\texttt{BRIDGE}\xspace}

\newcommand{\jax}[0]{\texttt{JAX}\xspace}

\newcommand{\tone}[0]{\texttt{TEST1}\xspace}
\newcommand{\ttwo}[0]{\texttt{TEST2}\xspace}
\newcommand{\tthree}[0]{\texttt{TEST3}\xspace}

\begin{document} 

   \title{Differentiable fuzzy cosmic web  for field-level inference}

\author{P. Rosselló\inst{1}\inst{2}\fnmsep\thanks{pere.rossello@iac.es} \and F.-S. Kitaura
\inst{1}\inst{2} \fnmsep\thanks{fkitaura@ull.edu.es}
\and D. Forero Sánchez \inst{3,4} \and F. Sinigaglia \inst{5}\inst{6} \inst{1}\inst{2} \and G. Favole \inst{1}\inst{2}
}

\institute{
Instituto de Astrof\'{\i}sica de Canarias, s/n, E-38205, La Laguna, Tenerife, Spain  
\and 
Departamento de Astrof\'{\i}sica, Universidad de La Laguna, E-38206, La Laguna, Tenerife, Spain
\and 
Departament de Física Quàntica i Astrofísica, Universitat de Barcelona, Martí i Franquès 1, E08028 Barcelona, Spain
\and
Institut de Ciències del Cosmos (ICCUB), Universitat de Barcelona (UB), c. Martí i Franquès, 1, 08028 Barcelona, Spain
\and
Département d’Astronomie, Université de Genève, Chemin Pegasi 51, CH-1290 Versoix, Switzerland
\and 
Institut für Astrophysik, Universität Zürich, Winterthurerstrasse 190, CH-8057 Zürich, Switzerland
}

  \date{Received XX; accepted XX}

  \abstract
   {A comprehensive analysis of the cosmological large-scale structure derived from galaxy surveys involves field-level inference, which requires a forward-modelling framework that simultaneously accounts for structure formation and tracer bias.}
   {While structure formation models are well understood, the development of an effective field-level bias model remains{ challenging}, particularly in the context of tracer  perturbation theory within Bayesian reconstruction methods, which we address in this work.}
   {To bridge this gap, we developed a differentiable model that integrates augmented Lagrangian perturbation theory and non-linear, non-local, and stochastic biasing. At the core of our approach is the Hierarchical Cosmic-Web Biasing Nonlocal (HICOBIAN) model, which provides a{ description of a field with a positive number of tracers} while incorporating a long- and short-range non-local framework via cosmic-web regions and deviations from Poissonity in the likelihood. A key insight of our model is that transitions between cosmic-web regions are inherently smooth, which we implemented using sigmoid-based gradient operations. This enables a fuzzy and, hence, differentiable hierarchical cosmic-web description, making the model well-suited for machine-learning frameworks.}
   {We tested the practical implementation of this model through GPU-accelerated computations implemented in \jax, the \wonj code, enabling a scalable evaluation of complex biasing. Our approach accurately reproduces the primordial density field within associated error bars derived from Bayesian posterior sampling within a self-specified setting (meaning that inference is performed on data generated by the exact same forward model) as validated by two- and three-point statistics in Fourier space. Furthermore, we demonstrate that the methodology approaches the maximum encoded information consistent with Poisson noise. 
   We also demonstrate that the bias parameters of a higher order non-local-bias model can be accurately reconstructed within the Bayesian framework{ for bias models with eight parameters}.
   }
   {We introduce a Bayesian field-level inference algorithm that leverages the same forward-modelling framework used in galaxy, quasar, and Lyman-alpha-forest mock-catalogue generation -- including non-linear, non-local and stochastic bias with redshift space distortions -- providing a unified and consistent approach to the analysis of large-scale cosmic structure.}


\maketitle

\section{Introduction}

{ The primordial density fluctuations provide the initial conditions for structure formation and encode statistical signatures of the early Universe}. These early inhomogeneities, seeded during the initial moments after the Big Bang, evolved under gravity to form the vast cosmic-web structure we observe today. Reconstructing the initial conditions from present-day observations of luminous tracers -- such as galaxies, quasars, Lyman-$\alpha$ forests, intensity maps, or from the 21cm line -- offers a powerful probe of fundamental physics, including inflation, dark matter, dark energy, and gravity itself. The scientific community is making significant efforts to map the matter distribution in the Universe, as exemplified by current large-scale spectroscopic surveys such as the Dark Energy Spectroscopic Instrument (\textit{DESI})  \citep[][]{DESI,DESI_DR1},  \textit{Euclid} \citep[][]{Euclid,EUCLID_Q1}; and future ones such as the Prime Focus Spectrograph (\textit{PFS}) \citep[][]{Takada_2014}, MUltiplexed Survey Telescope (\textit{MUST}) \citep[][]{MUST_2024}, and the Nancy Grace Roman Space Telescope (\textit{Roman}) \citep[][]{Wang_2022}. 

The first attempts to recover the primordial density fluctuations from the galaxy distribution were based on inverse reconstruction methods \citep[e.g.][]{Bertschinger_1989,Nusser_1992,Monaco_1999}. These approaches proved to be highly effective for sharpening features such as baryon acoustic oscillations (BAOs; \citealt{Eisenstein_2007}), and were even proposed as tools to constrain primordial non-Gaussianities \citep{Shirasaki_2021}.
However, the accuracy of these inverse techniques is fundamentally limited by the non-invertibility of the gravitational evolution in the non-linear regime in the absence of full phase-space information, particularly due to shell-crossing. Once multiple streams of matter overlap, information about the initial conditions is irreversibly lost, 
as the full phase-space dynamics -- crucial for a unique reconstruction -- are no longer accessible through the evolved tracer distribution alone.

To overcome the limitations of inverse methods, forward modelling was introduced as a principled alternative. Instead of inverting the non-linear structure formation process, forward approaches start from initial conditions and evolve them through a physical model of gravitational dynamics to generate predictions for present-day observables \citep[][]{Kitaura_2013a,Jasche_2013,Wang_2013}.  The first application of a forward-modelling reconstruction (including peculiar motions) to observational data was presented by \citet{Kitaura_2012a}. All these works demonstrated the viability of sampling high-dimensional posterior distributions consistent with both data and gravitational physics, laying the foundation for later advances in bias modelling and inference methods. Subsequent developments considered improved mask handling \citep[][]{jaschePresentCosmicStructure2015, lavauxUnmaskingMaskedUniverse2016,Ata_2021}, improved gravity solvers using particle-mesh codes \citep[][]{Wang_2014,Jasche_2019,Horowitz_2019}, and emulators \citep[][]{doeserBayesianInferenceInitial2024}. Further extensions incorporated light-cone evolution \citep[][]{Kitaura_2021, lavauxSystematicfreeInferenceCosmic2019}, primordial non-Gaussianities \citep[][]{andrewsBayesianFieldlevelInference2023}, effective field theory (EFT) approaches \citep[][]{stadlerCosmologyInferenceField2023}, and hydrodynamics \citep[][]{horowitzDifferentiableCosmologicalHydrodynamics2025}. 
These techniques have also been applied to alternative tracers, such as the Lyman-$\alpha$ forest \citep[e.g.][]{Kitaura_2012,Horowitz_2019b,porqueresInferringHighredshiftLargescale2019,Horowitz_2021,Horowitz_2022}.{ A benchmark of computational efficiency in this context is presented in \citet[][]{Bayer2023}.}

Bayesian statistics offers the ideal framework for field-level inference, enabling clear specification of prior knowledge and observational uncertainties \citep[][]{Kitaura_2008}. In this setting, reconstruction becomes a matter of sampling from the posterior distribution of initial conditions, informed by both the underlying physics and the observed distribution of tracers.

This approach, however, faces a significant computational barrier. High-fidelity dark-matter simulations capable of resolving the halos that host galaxies require substantial computational resources -- often hundreds of thousands to millions of CPU hours -- rendering brute-force posterior sampling impractical \cite[see state-of-the-art $N$-body simulations, e.g.][]{Garrison_2018,Chuang_2019,Ishiyama_2021}.

To mitigate this challenge, Lagrangian perturbation theory (LPT)  offers a computationally efficient alternative to full gravity solvers \citep[][]{Bernardeau_2002}, such as $N$-body simulations \citep[][]{Angulo_2022}. While being approximate, LPT accurately captures the gravitational evolution of matter on large, quasi-linear scales and serves as an effective forward model.
Additional advancements have further extended and  regularised these approximations, enabling reliable predictions down to megaparsec -- or even sub-megaparsec -- scales \citep[][]{Kitaura_2013,Kitaura_2024}.  

This strategy is particularly effective when the forward model is constrained to its domain of validity and coupled to the observed tracers through an effective field-level bias prescription. The core challenge then lies in constructing a robust and physically motivated bias model. A critical first step is to ensure that large-scale bias is accurately captured, which necessitates the inclusion of non-linear and non-local contributions up to the third order \citep[][]{McDonald_2009}.

A longstanding limitation of perturbation-theory-based bias models is that truncation at a fixed order often introduces unphysical behaviour \citep[][]{McDonald_2009,Schmittfull_2019,Werner_2020}. Specifically, these models can yield non-positive densities and exhibit oscillatory or noisy behaviour when extended into the highly non-linear regime. However, recognising that non-local-bias terms at different orders correspond to distinct morphological features of the cosmic web opens the door to a more stable and flexible framework \citep[][]{Kitaura_2022}. For each morphological feature, we can then assume a local non-linear-bias model tailored to the specific properties of that structure. In this way, the classification of the cosmic web into distinct patterns effectively acts as a diagonalising operation, isolating the dominant local contributions to the tracer-matter relationship within each structural environment. This enables the systematic inclusion of higher order terms while preserving physical consistency and numerical stability down to small scales, making the framework well-suited for precision field-level inference.
This ansatz provides a general and flexible framework applicable to a wide range of tracer populations and has demonstrated unprecedented accuracy in field-level bias modelling of point-like tracers such as dark-matter halos \citep[][]{Balaguera_2023,Coloma_2024} and the baryonic components resolved in cosmological hydrodynamical simulations -- including ionised gas, neutral hydrogen, and the Lyman-$\alpha$ forest \citep[][]{Sinigaglia_2021,Sinigaglia_2022,Sinigaglia_2024,Sinigaglia_2024b}.

In this paper, we introduce \wonj\footnote{Bayesian Reconstruction and Inference of Data-driven Generative Environments: github.com/pererossello/bridge-repo.}, a differentiable, GPU-accelerated framework for field-level inference written in \jax \citep[][]{jax2018github}. In Sect.~\ref{sec:methods}, we detail the technical implementation of the structure formation and bias models. Section \ref{sec:results} presents a series of computational tests demonstrating that the code successfully recovers both the primordial density field and the bias parameters of a long-range non-local-bias model at resolutions of 5 and 10 \mpch. Finally, we describe how we applied \wonj using a combined long- and short-range bias model, showing that the framework can robustly recover the primordial density field even in the presence of highly complex tracer bias.

\begin{figure*}[]
\centering
\includegraphics[width=0.9\textwidth]{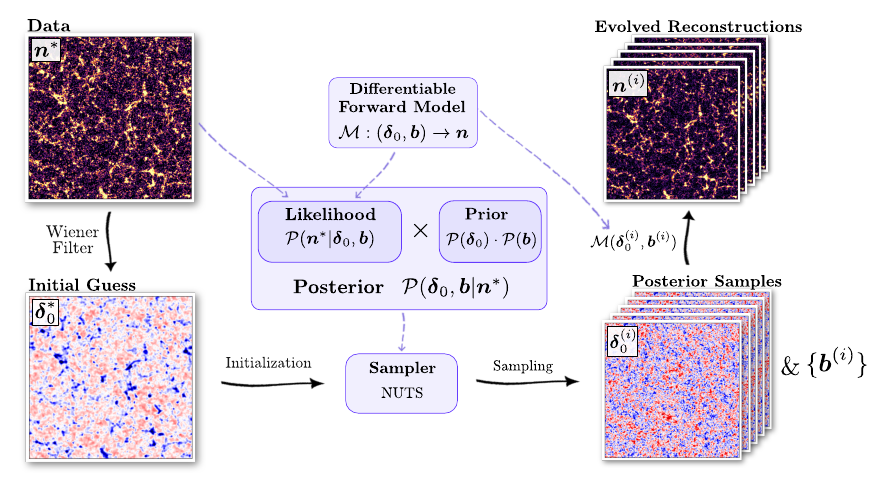}
    \caption{Schematic overview of the \wonj pipeline. The framework combines a scalable, differentiable structure-formation model with a flexible field-level bias prescription, all implemented in a GPU-accelerated \jax environment. This enables efficient Bayesian inference of the primordial density field from tracer observations, with support for complex, non-local-bias models and multi-resolution analysis. The modular design allows for the seamless integration of physical models while maintaining end-to-end differentiability and high computational performance.
    }
    \label{fig:webonjax}
\end{figure*}

\section{Methods}
\label{sec:methods}

In this section we present the technical details of the \wonj code. A flowchart showing the architecture of \wonj is presented in Fig.~\ref{fig:webonjax}.

\subsection{Field-level inference}

    We first consider a galaxy distribution, mapped onto a comoving volume by converting redshifts to distances, as the observed tracer field. 
    With
    \(
    \vect{n}^*\in\mathbb{N}^{N^{3}},
    \)
    we denote the observed tracer intensity  on a
    cubic, comoving grid consisting of \(N^{3}\) cells.  
    A differentiable forward model, 
    \(
    \mathcal{M,}\) maps a white-noise realisation, \( \vect{\nu} \), 
    which defines the initial density field through a linear power spectrum, together with 
    tracer bias parameters \(\vect{b}\)  and  cosmological parameters   \(\vect{\Omega}\) to the expected tracer field \(\bar{\vect{n}}\): 
    \begin{equation}
    \mathcal{M}\!:\!(\vect{\nu},\vect{b}, \vect{\Omega})\mapsto{\bar{\vect{n}}} \ .
    \end{equation}
    The forward model, \(\mathcal{M,}\) is constructed as a composition of three sequential maps,
    \(
    \mathcal{M} = \mathcal{M}_\text{b} \circ \mathcal{M}_{\Psi} \circ \mathcal{M}_\delta,
    \)
    where \(\mathcal{M}_\delta\) is a linear transformation that maps
a white-noise realisation, \(\vect{\nu,}\) to the initial overdensity field,
\(\vect{\delta}_0,\) at high redshift using the linear matter power spectrum
(see Appendix \ref{app:white} for details on the input field parametrisation); \(\mathcal{M}_{\Psi}\) is the gravity solver that evolves the
initial field, \(\vect{\delta}_0,\) into the non-linear dark-matter over-density
field, \(\vect{\delta,}\) at a specified redshift using an efficient gravity
solver; and \(\mathcal{M}_\text{b}\) maps the evolved matter field, \(\vect{\delta}\),
together with the bias parameters, \(\vect{b}\), to the mean tracer field,
\(\bar{\vect{n}}\), employing a cosmic-web-dependent, non-local-bias model.  

    For each cell, we assume the data arise from an independent
    discrete distribution with mean \(\bar n_i\), which may include additional
    parameters, \(\vect{p,}\) modelling survey completeness or super-Poissonian dispersion. The likelihood, which incorporates the forward model, reads 
    \begin{equation}
    \mathcal{L}(\vect{\nu}, \vect{b}, \vect{p},\vect{\Omega})
          \equiv
          \mathcal{P}(\vect{n}^* | \vect{\nu}, \vect{b}, \vect{p},\vect{\Omega}) 
          \, ,    \label{eq:likelihood}\end{equation}
    where $\mathcal{P}$ denotes the probability of observing $\vect{n}^*$ given the model parameters.
    
    Performing Bayesian inference at the field level amounts to sampling from the posterior distribution of the latent initial conditions and model parameters, given the observed tracer field $\vect{n}^*$. This posterior is given by
    \begin{equation}
    \mathcal{P}(\vect{\nu}, \vect{b}, \vect{p}, \vect{\Omega} \mid \vect{n}^*) \propto
    \mathcal{P}(\vect{\nu})\,
    \mathcal{P}(\vect{b}, \vect{p}, \vect{\Omega})\, \mathcal{L}(\vect{\nu}, \vect{b}, \vect{p}, \vect{\Omega}) \,,
    \label{eq:posterior}
    \end{equation}
    where $\mathcal{P}(\vect{\nu})$ encodes prior information on the initial conditions, and $\mathcal{P}(\vect{b},\vect{p},\vect{\Omega})$ represents prior knowledge on the bias and cosmological parameters. The forward model was constructed such that the latent field, \(\vect{\nu,}\) is whitened, and therefore \(\mathcal{P}(\vect{\nu})\) is an \(N^3\)-dimensional multi-variate Gaussian with zero mean and identity covariance.  The bias and additional likelihood model parameters, \(\vect{b}\) and \(\vect{p}\), which in our case were strictly positive, were assigned independent log-normal priors. The probability density function of the log-normal distribution{ for the bias parameters implemented in the \wonj code} is given by
    \begin{equation}
         \boxed{ f(b;\, \mu, \sigma) = 
          \frac{1}{b \sigma \sqrt{2\pi}} \exp\left[ -\frac{(\ln b - \mu)^2}{2\sigma^2} \right] \, , \quad  b > 0 }\, ,
          \label{eq:lognormal}
    \end{equation}
    where \(\mu\) and \(\sigma\) are fixed hyperparameters that define the mean and standard deviation of \(\ln b\).
Throughout the remainder of this work, we fixed the cosmological parameters, \(\vect{\Omega,}\) to their fiducial values and omitted their explicit dependence for notational simplicity.

\subsection{Posterior sampling}

We exploited the differentiability of the forward model to efficiently sample the high-dimensional posterior using gradient-based Hamiltonian Monte Carlo (HMC) sampling \citep[][]{Duane_1987}. 
HMC was first applied to Bayesian large-scale-structure inference by \citet{Jasche_2010} and subsequently to observational data by \citet{Jasche_2010b}.

 { We implemented within the \wonj code various posterior sampling methods, such as  second- and fourth-order HMC sampling and the \citet{Blanes_2014} HMC  variant. The numerical tests shown in this work were finally performed with the publicly available} No-U-Turn Sampler (NUTS) in
\texttt{NumPyro} \citep[][]{Hoffman_2011, phanComposableEffectsFlexible2019}.   
During the adaptation (burn-in) phase, the algorithm tunes the step size,
trajectory length, and the mass matrix, which is kept diagonal due to memory constraints. During sampling, we fixed the adapted mass matrix and step size and integrated the trajectories{ with a NUTS tree depth equal to 10.}  

To accelerate convergence, each chain was initialised from a deterministic state
obtained by Wiener-filtering the observed tracer density contrast \citep[][]{Zaroubi_1995},
\(\vect{\delta}^*_\text{tr} \equiv \vect{n}^* N^3 / N_\text{tr} - 1\), 
back to an approximate initial density field, \(\vect{\delta}^{(0)}_{0}\),
which was transformed into white noise,
\(\vect{\nu}^{(0)}\); {  $N_{\rm tr}$ is the total number of tracers}. { In this first step, we assumed the linear power spectrum to be known and ignored peculiar motions.}

\subsection{Gravity solver}

We modelled large-scale-structure (LSS) formation with Augmented Lagrangian perturbation theory (ALPT; \citealt{Kitaura_2013}), which merges second-order LPT (2LPT; \citealt{Zeldovich_1970,Buchert_1994,Bouchet_1995,Catelan_1995}) for long-range tidal displacements and a spherical-collapse (SC) solution \citep{Bernardeau_1994,Neyrinck_2016} for short-range dynamics. This was achieved by decomposing the particle displacement field as{ implemented in the \wonj code:} 
\begin{equation}
    \boxed{
    \vect{\Psi} = \vect{\Psi}_{\rm L}+\vect{\Psi}_{\rm S}}\,,
\end{equation}

\noindent with $\vect{\Psi}_{\rm L}=\mathcal{K}(\vect{q},r_s)\circ\vect{\Psi}_{\mathrm{2LPT,}}$  $\vect{\Psi}_{\rm S}=(1-\mathcal{K})\circ\vect{\Psi}_{\mathrm{SC}}$, and $\mathcal{K}$ a smoothing kernel of fixed radius, $r_s = 4$ \mpch,{} following \citet{Kitaura_2013}. In future work, we plan to sample this parameter within the Bayesian framework and consider nLPT for the long-range interaction.

This hybrid formulation retains analytic dependence on cosmological parameters while suppressing shell crossing. Coupled with empirical bias schemes such as the Bias Assignment Method (BAM).  \citep{Balaguera_2018,Balaguera_2020}, ALPT reproduces $N$-body halo statistics down to $k\lesssim0.4$\,\hmpc{}, the scale at which most cosmological information resides \citep{Hahn_2021}. An even higher accuracy was achieved with the HICOBIAN bias model considered in this work (see Sect.~\ref{sec:nonlocal}).

We note that ALPT is substantially more efficient in terms of both computational speed and memory usage compared to fast particle mesh (PM) methods \citep[e.g.][]{Tassev_2013,Feng_2016,Klypin_2018}, as shown in the comparison work by \citet{Blot_2019}, making it especially advantageous for scalable Bayesian inference frameworks.  Moreover, ALPT can be extended via iteration (eALPT) to reach sub-megaparsec accuracy \citep{Kitaura_2024} by indirectly emulating the effect of a viscous stress tensor in the equations of motion, thereby capturing aspects of non-linear dynamics such as vorticity. 
Recent GPU-based PM methods have shown noteworthy advancements \citep[e.g.][]{Modi_2021,Li_2024}. Alternative promising approaches to approximate structure formation are offered by emulators trained on large ensembles of simulations \citep[][]{Kodi_2020,Conceiccao_2024}. However, for the time being, we opt to rely on analytic solutions with explicit cosmological dependence.

\subsection{Bias model framework}

When employing an approximate gravity solver, the evolved density field should not be directly trusted at the particle level. Instead, it is only reliable on a coarse mesh with a resolution of a few megaparsecs, where the approximation remains accurate. As a result, the bias relation between matter and tracers is no longer deterministic -- as in approaches that apply a halo finder directly to particle distributions -- it must be modelled as an effective field-level bias \citep[][]{Schmittfull_2019}.
In this framework, the luminous tracers of LSS exhibit a complex, non-linear, non-local, and stochastic relationship with the underlying matter field \citep[][]{McDonald_2009,Desjacques_2018}. Accurately capturing this relationship is essential for unbiased inference and requires flexible and physically motivated modelling approaches that operate at the field level.

We took advantage of a positive, non-local-bias modelling framework that employs local-bias expansions within distinct morphological regions of the cosmic web \citep[][]{Kitaura_2022,Coloma_2024,Sinigaglia_2024}. 

\subsection{Deterministic local-bias model}

The bias relation between tracers and the underlying matter field has long been known to be non-linear. One of the earliest approaches involved a local perturbative expansion of the tracer overdensity, introduced by \citet[][]{Fry_1993}:
\begin{equation}
\delta_{\rm tr} = \sum_{m=0}^\infty \frac{b_m}{m!} \delta^m\,,
\end{equation}
where \(\delta_{\rm tr}\) denotes the tracer overdensity, \(\delta\) the matter overdensity, and \(b_m\) the bias coefficients. While conceptually straightforward, this expansion suffers from the drawback that it may yield unphysical negative tracer densities when truncated at any order.

To address this issue, an alternative logarithmic formulation was proposed around the same time by \citet{Cen_1993}:
\begin{equation}
\log(1 + \delta_{\rm tr}) = \sum_{m=0}^\infty \frac{c_m}{m!} \left[\log(1 + \delta)\right]^m\,,
\end{equation}
which ensures positivity of the tracer field by construction. To a linear order, this expansion corresponds to a power-law bias model, which has been particularly effective in modelling the Lyman-$\alpha$ forest. In this context, it leads to the fluctuating Gunn--Peterson approximation, where the optical depth, \(\tau,\) is related to the matter density field by 
$\tau \propto (1 + \delta)^\alpha$ \citep[][]{Bi_1997}. An analogous relationship was found for the functional dependency between the molecular gas temperature and dark-matter density \citep[][]{Hui_1997}.

Interestingly, a power-law bias model in Eulerian space can be related to a linear bias in Lagrangian space -- i.e. in the coordinate system of the initial conditions -- according to the continuity equation, which leads to a log-normal evolution of the density field \citep[][]{Coles_1991}. In this picture, a linear Lagrangian bias naturally evolves into a power-law bias at later times.

In the context of galaxy bias, the polynomial perturbative expansion has historically been preferred. A foundational physical interpretation of bias was introduced even earlier by \citet{Kaiser_1984} through the excursion set model, where galaxies preferentially form in high-density peaks of the  matter distribution.
Building on this idea, \citet{Kitaura_2014} proposed a hybrid model that combines the power-law bias of \citet{Cen_1993} with the threshold bias of \citet{Kaiser_1984} to model galaxy number counts as a function of the underlying dark-matter field. This framework  was later refined by \citet{Neyrinck_2014}, the authors of which replaced the hard threshold with an exponential cutoff. The resulting model captures the average bias behaviour of halos with remarkable accuracy, particularly in low-density environments. 

Although the power-law and threshold bias models are degenerate with respect to two-point statistics -- both amplifying power on large scales -- this degeneracy can be broken using three-point statistics, which provide a means to distinguish between different biasing mechanisms (see also \citealt{Kitaura_2015} for the definitions of three-point statistics calculations used in this work). The hybrid model has proven especially useful in this regard and was employed to generate thousands of mock catalogues of luminous red galaxies for the final BOSS data release \citep[see the PATCHY mocks in][]{Kitaura_2016}.

To date, the suppression of tracer densities has primarily focused on low-density regions, following the picture introduced by Kaiser. However, another important effect in the context of halo and galaxy formation is halo exclusion \citep[][]{Baldauf_2013}, which accounts for the saturation of tracer objects in highly overdense environments. This mechanism introduces a natural suppression of tracer abundance in high-density regions, complementing the low-density suppression from the original threshold bias picture \citep[][]{Coloma_2024}. A similar concept has also been applied in modelling the Lyman-$\alpha$ forest, where saturation effects in flux absorption become significant in dense regions  \citep[][]{Sinigaglia_2024}.

Motivated by these considerations, we modelled { and implemented} the expected tracer number density in the \wonj
code using a power-law component modulated by both a low-pass and a high-pass filter:
\begin{equation}    \boxed{  \bar{n}_i =
      C
      \left(1 + \delta_i\right)^{\alpha}
      \exp \left[-
      \left(
      \frac{1 + \delta_i}{\rho_\text{}}
      \right)^{ \epsilon_{\text{}}}
      \right]
      \exp \left[-
      \left(
      \frac{1 + \delta_i}{\rho'_\text{}}
      \right)^{ -\epsilon'_{\text{}}}
      \right]} \ ,
      \label{eq:local_bias}
\end{equation}
where \(C\) is a normalisation constant defined such that \(\sum_i \bar{n}_i = \sum_i n^*_i \equiv  N_\text{tr}\), where \(N_\text{tr}\) is the total number of tracers in the data. All parameters are positive.
 We assume that the observed tracer count in each cell is a stochastic realisation drawn from a discrete probability distribution, \(\mathcal{P}^{\text{tr}}_i\), characterised by a mean expected value, \(\bar{n}_i\), provided by the forward model.

\subsection{Stochastic bias model}

The coarse resolution inherent to the field-level framework necessitates the introduction of a stochastic component to the bias model \citep[][]{Dekel_1999,Sheth_1999}, which is not explicitly present in deterministic high-resolution $N$-body simulations. Nonetheless, the lack of full phase-space information after shell-crossing also introduces an uncertainty that requires a probabilistic treatment in the likelihood.

The most basic discretisation assumes a Poisson likelihood, where the variance equals the mean tracer count. 
The complete integration of this likelihood beyond the second moment within the Bayesian inference framework has been studied for a long time \citep[][]{Kitaura_2008,Kitaura_2010}. However, this assumption holds only in very limited regimes -- typically for low tracer densities and in relatively homogeneous environments.
As originally predicted by \citet{Peebles_1980}, (anti-)correlations between tracers below the mesh resolution introduce deviations from pure Poisson statistics. These unresolved sub-grid correlations give rise to either sub- or super-Poissonian noise characteristics. To account for this, various probability distribution functions (PDFs) have been proposed in the literature \citep[][]{Saslaw_1984,Sheth_1995}, most of which introduce additional degrees of freedom to model the over-dispersion  commonly observed in galaxy and halo distributions relative to the Poisson expectation \citep[][]{Somerville_2001, Casas_Miranda_2002}. \citet{Pellejero_2020} demonstrated that the negative binomial distribution can be parametrised to reproduce a wide range of functional dependencies in the over-dispersed deviations from Poisson statistics.

We implemented the  negative binomial distribution to model \(\mathcal{P}^{\text{tr}}_i\), thereby accounting for potential over-dispersion in the tracer counts. This distribution was first considered in the context of field-level bias in \citet{Kitaura_2014} and later implemented in the Bayesian framework in \citet{Ata_2015}. Our implementation of the negative binomial distribution in this work relies on the  gamma--Poisson mixture modelling, in which the Poisson rate, \(\lambda_i,\) is treated as a random variable drawn from a gamma distribution. {The corresponding} hierarchical formulation { implemented in the \wonj code} is given by
\begin{eqnarray}
     \boxed{ \lambda_i \sim  \text{Gamma}(\beta, \bar{n}_i / \beta)} && \\
      \boxed{n^*_i    \sim \text{Poisson} (\lambda_i)}&& \,,
\end{eqnarray}
where \(\bar{n}_i\) denotes the expected tracer count in cell \(i\), and \(\beta\) is the over-dispersion parameter. In the limit \(\beta \rightarrow \infty\), the gamma distribution becomes sharply peaked, recovering the standard Poisson likelihood. 
{In the gamma distribution, the first argument is the shape parameter and the second argument is the scale parameter.}
With this model, the (negative log) likelihood reads
\begin{equation}
      - \log \mathcal{L}(\vect{\nu}, \vect{b}, \beta)
      \equiv
      \sum_{i=1}^{N^3}
      - \log \mathcal{P}^{\text{tr}}_i(n^*_i | \mathcal{M}(\vect{\nu}, \vect{b}), \beta)
      \ ,
      \label{eq:log_likelihood}
\end{equation}
with $\vect{b} = (\alpha, \rho, \epsilon, \rho^\prime, \epsilon^\prime)$ from Eq.~\ref{eq:local_bias}.
While sub-Poissonian behaviour is indirectly captured by the high-density damping exponential factor, an explicit treatment can be implemented using the Conway--Maxwell--Poisson distribution \citep[][]{Daly2015} or extensions of the binomial  distribution  \citep[][]{2024MNRAS.530.3458V}.

\subsection{Non-local bias and cosmic-web dependence}
le\label{sec:nonlocal}

In the previous sections, we introduced a local non-linear-bias model to describe the average bias relation and subsequently incorporated scatter to account for stochasticity. However, this stochastic component should be understood as a feature of the coarse-grained nature of the effective bias approach, without  superseding  deterministic contributions from non-local-bias effects. 

Galaxy bias is shaped not only by the local matter density, but also by the surrounding large-scale tidal environment and the small-scale curvature of the density field. These non-local dependencies stem from the intrinsic nature of gravitational collapse and galaxy formation. A general framework for incorporating such effects into bias expansions was introduced by \citet{McDonald_2009}.  In another particularly relevant study, \citet{Chan_2012} connected long-range tidal tensor invariants to higher order bias terms. \citet{Kitaura_2022} later related the invariants of the Hessian of a field to the morphological classifications of the cosmic web.

These developments motivated the extension of bias models to explicitly include cosmic-web morphology (see Appendix~\ref{app:types}). Galaxies inhabiting different large-scale structures, including voids, sheets, filaments, and knots, exhibit systematically different biasing behaviours, as shown in simulations and observations \citep[e.g.][]{Nuza_2014,Filho_2015}.

Several approaches exist for incorporating cosmic-web morphology. One method involves binning tidal field tensor invariants, which can result in a high-dimensional classification if multiple invariants and their dependence on local density are included \citep[][]{Kitaura_2022}. Alternatively, one can compress this information into four characteristic environments, sacrificing some modelling granularity for interpretability -- this leads to the so-called $\Phi$-web classification, based on the large-scale tidal field \citep[][]{Hahn_2007}.

To recover finer biasing distinctions, we introduced a hierarchical structure by embedding a second classification based on the Hessian of the density field (the $\delta$-web) within each $\Phi$-web region. This short-range non-local bias captures small-scale tidal effects rooted in tidal torque theory \citep[][]{Heavens_1988}. The result is the HICOBIAN model, which partitions the space into 16 regions \citep[][]{Coloma_2024}. For each region, it applies a different local-bias relation, which is determined by the specific bias parameters of model Eq.~\ref{eq:local_bias}.  This approach gains physical insight with computational efficiency and outperforms fine binning of the tidal field in terms of both precision and interpretability, as both short- and long-range non-local biases are incorporated.

Given that gravitational evolution is inherently non-local, the shortcomings of ALPT relative to full gravity solvers can be effectively absorbed into non-local-bias terms. This explains the excellent performance of more recent studies employing ALPT \citep[e.g.][]{Balaguera_2023,Coloma_2024}. This insight provides a foundation for extending the bias framework to capture deviations arising from alternative gravity models \citep[][]{Garcia-Farieta_2024}, where characteristic bias parameters and higher order statistics including Legendre multipole expansions would serve as indicators of such modified dynamics.

In the context of this work, it is important to note that most cosmic-web classifiers rely on hard, non-differentiable thresholds (e.g. eigenvalue sign changes), which are incompatible with gradient-based inference techniques such as HMC sampling. To address this, we adopted a fuzzy, differentiable classification scheme using sigmoid-based transitions. This enables smooth, physically motivated boundaries between web types and integrates seamlessly with our differentiable, field-level inference framework.

\begin{figure*}[]
\centering
\includegraphics[width=0.7\textwidth]{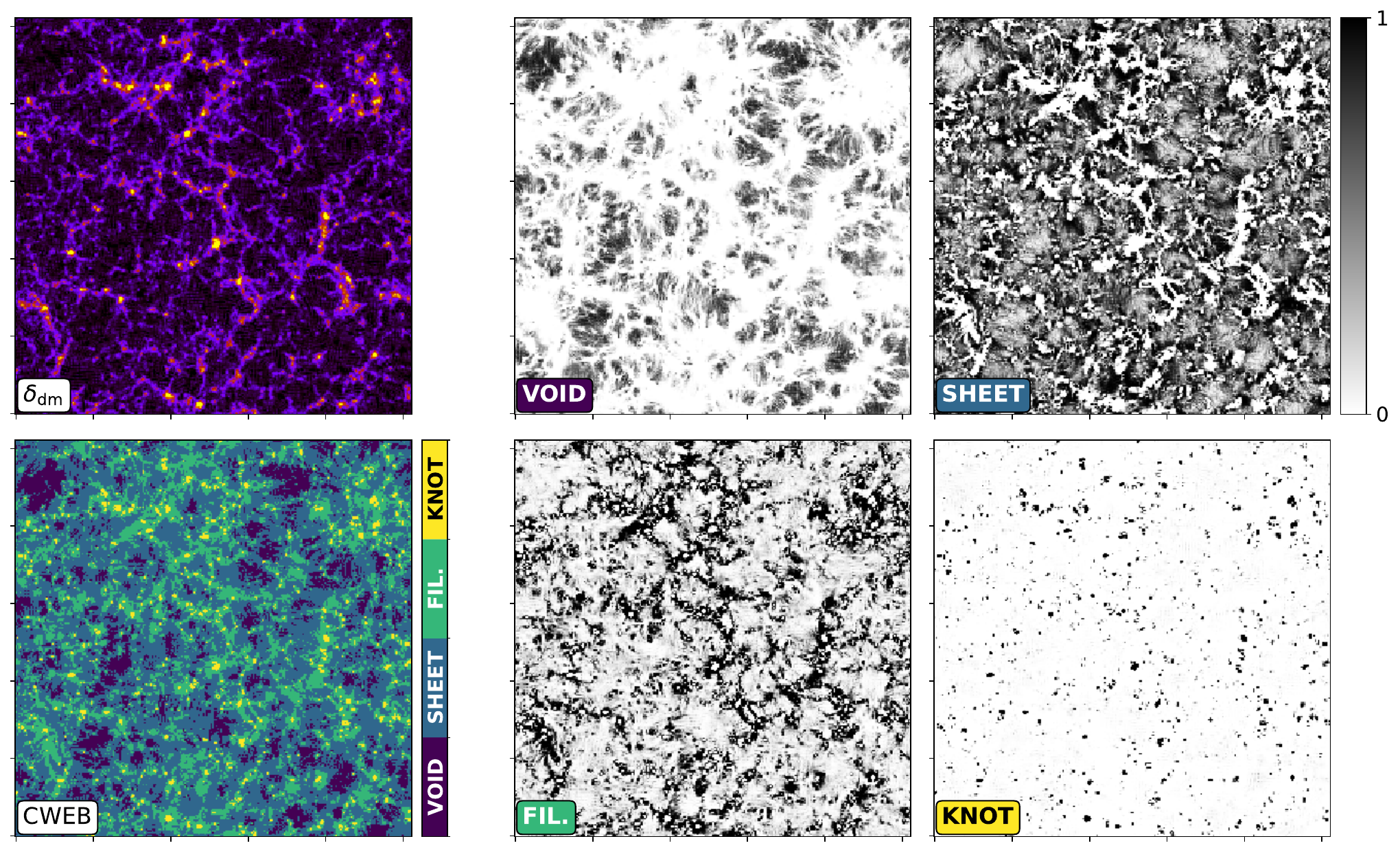}
    \caption{
        Example of fuzzy cosmic-web classification with \(N=256\) and $\Delta L=1.7\,h^{-1}\mathrm{Mpc}$. Top left: Evolved dark-matter density contrast field. Bottom left: `Hard' $\Phi$-web classification. Right panels: Fuzzy membership weights, $p_i^{(\mathrm{V})}$, $p_i^{(\mathrm{S})}$, $p_i^{(\mathrm{F})}$, and $p_i^{(\mathrm{K})}$, for voids, sheets, filaments, and knots, respectively, as defined in Eq.~\ref{eq:soft_cweb_p}.
    }
    \label{fig:cweb}
\end{figure*}

\subsection{Differentiable fuzzy cosmic-web classification}

In the top-down structure-formation scenario proposed by \citet{Zeldovich_1970}, cosmic structures form via anisotropic collapse governed by the tidal field tensor. The process unfolds along the tensor’s eigenvectors, with collapse first occurring along the direction associated with the largest eigenvalue -- a phenomenon known as pancake formation.
    The term cosmic web was coined by \citet[][]{Bond_1996nat} to describe the large-scale structure of the Universe, characterised by the emergence of distinct morphological components such as filaments, walls, and nodes arising from anisotropic gravitational collapse. This interpretation was supported by shape parameters such as ellipticity and prolateness, derived from the eigenvalues of the deformation tensor and linked to the Zel'dovich approximation.
    A more systematic and operational classification of the cosmic web was later proposed by \citet{Hahn_2007}, the authors of which used the eigenvalues of the tidal (equal to the velocity shear within the Zel'dovich approximation) tensor to categorise regions of space into knots, filaments, sheets, and voids, depending on the number of collapsing directions. This method provided a quantitative tool to identify web environments in cosmological simulations.
    
While widely used, cosmic-web classification schemes remained largely phenomenological, lacking a firm quantitative foundation. A more rigorous interpretation began to emerge with the connection to the framework of non-local bias, which provided a physically motivated context for understanding how large-scale structure influences halo and galaxy formation -- an approach we adopted in this work (see Appendix \ref{app:types}).

We now consider a cosmic-web classification scheme based on the tidal field tensor,
\(\partial_{j} \partial_{k} \Phi\)
, where $\Phi = \nabla^{-2} \delta$ is the gravitational potential. Let
\(\lambda_i^{(1)} < \lambda_i^{(2)} < \lambda_i^{(3)}\) be the eigenvalues of the tidal field tensor at the \(i\)th cell. The cosmic-web classification at the \(i\)th cell can be compactly expressed as
\begin{equation}
      W_i =
      \sum_{n=1}^{3}
      h (\lambda_i^{(n)} - \lambda_\text{th})
      \ ,
      \label{eq:hard_cweb}
\end{equation}
where \(h\) is the Heaviside step function
and \(\lambda_\text{th}\) is a free parameter.  In this work, we set \(\lambda_\text{th}=0.05\) (see \citealt{Forero_2009} for the introduction of a threshold in the cosmic-web classification). For recent studies regarding the choice of \(\lambda_\text{th}\), see \citet{Coloma_2024} and \citet{olexUniversalPhysicallyMotivated2025}. 
\(W_i\) counts the number of eigenvalues above the threshold; the four resulting values of
0, 1, 2, and 3 correspond to voids, sheets, filaments, and knots, respectively. This cosmic-web definition is problematic if introduced in a forward model that aims to be differentiable with respect to \(\vect{\nu}\). This is because the cosmic-web classification mesh, \(\vect{W,}\) depends on \(\vect{\nu}\) through the potential of \(\vect{\delta} = \mathcal{M}_{\Psi} \circ \mathcal{M}_{\delta} (\vect{\nu})\), but it does so in a manifestly non-differentiable way due to the presence of the Heaviside step function.
As a consequence, the gradients \(\partial W_i / \partial q_j\) are ill-defined, which precludes the use of gradient-based sampling methods.
To address this, we relied  { within the \wonj code}  on a fuzzy cosmic-web classification defined through the sigmoid weights, with
\begin{equation}\boxed{
      w^{(n)}_i =
      \sigma [s (\lambda^{(n)}_i -
            \lambda_\text{th})]}\,,
      \label{eq:sigmoid}
\end{equation}
where \(\sigma(x) = (1+\mathrm{e}^{-x})^{-1}\), and $s$ is the steepness parameter controlling how sharply \(w^{(n)}_i\) transitions around \(\lambda_{\text{th}}\). 
Because \(\sigma\) is strictly increasing and the eigenvalues are sorted, the weights inherit the order
\(w^{(1)} > w^{(2)} > w^{(3)}\).  
We converted these cumulative exceedance probabilities into mutually exclusive
`fuzzy' memberships for knots (K), filaments (F), sheets (S), and voids (V):
\begin{align}
      p^{(\text{K})}_i & = w^{(3)}_i \ ,                          \\
      p^{(\text{F})}_i & = w^{(2)}_i - w^{(3)}_i\ ,              \\
      p^{(\text{S})}_i & = w^{(1)}_i - w^{(2)}_i\ , \\
      p^{(\text{V})}_i & = 1- w^{(1)}_i   \ ,
      \label{eq:soft_cweb_p}
\end{align}
so that they represent the probabilities of having exactly three, two, one, or zero eigenvalues above the threshold, and, by construction, 
\(p^{(\mathrm V)}_i+p^{(\mathrm S)}_i+p^{(\mathrm F)}_i+p^{(\mathrm K)}_i=1\).
After introducing this smooth classification scheme, the forward model remained differentiable.
Regarding the steepness parameter \(s\), a larger value makes the fuzzy classification closer to the original classification, but it also risks causing numerical instabilities in the gradients. We therefore chose the largest \(s\) that preserves stable gradient computations. In Fig. \ref{fig:cweb}, we show a comparison of the `sharp' and smooth cosmic-web classification.

We now consider a cosmic-web-dependent bias model where each of the bias parameters from the local-bias model depend on the cosmic-web classification of the voxel. That is, given a bias parameter of \(b \in \{\alpha, \rho, \epsilon, \rho^\prime, \epsilon^\prime\}\), it now takes different values in each of the cosmic-web regions: \(b^{(\text{K})}, b^{(\text{F})}, b^{(\text{S})}, b^{(\text{V})}\). To ensure differentiability, we used the fuzzy cosmic-web classification, and the bias parameter at voxel \(i\) takes the value
\begin{equation}\boxed{
      b_i =
      b^{(\text{K})} p^{(\text{K})}_i +
      b^{(\text{F})} p^{(\text{F})}_i +
      b^{(\text{S})} p^{(\text{S})}_i +
      b^{(\text{V})} p^{(\text{V})}_i}
      \ .
\end{equation}

\subsection{Redshift-space-distortion modelling}

Redshift-space distortions (RSDs) arise when the observed position of a tracer deviates from its Hubble-flow distance due to peculiar velocities, making them sensitive probes of the growth rate of structure formation \citep[see][]{Kaiser_1987,Hamilton_1998}.
Redshift-space distortions are governed by two distinct types of peculiar motions: coherent large-scale flows, which arise from gravitational infall towards overdense regions, and small-scale random motions within quasi-virialised structures. For a variety of ways of modelling coherent flows, we refer the reader to \citet{Kitaura_2012c} and references therein \citep[see also][]{Wang_2012}.
There are several strategies for incorporating RSDs within a Bayesian framework (we chose the third), which we list below.
\begin{enumerate}
    \item Iterative sampling of the redshift-to-real-space mapping. 
    This approach iteratively reconstructs the real-space distribution from redshift-space observations using Gibbs sampling \citep[][]{Kitaura_2008,Kitaura_2012,Kitaura_2016a}. It is inspired by earlier iterative RSD correction methods \citep[see][]{Yahil_1991,Monaco_1999}.
    
    \item Forward modelling of RSDs at the tracer level.
    Originally introduced by \citet{Kitaura_2012a}, this method evolves tracers from sampled initial conditions to redshift space, where their final positions are compared with observed galaxy distributions to constrain the primordial density field. Small-scale virial motions can be addressed in two ways: either by collapsing elliptical groups of galaxies -- as proposed by \citet{Tegmark_2004} -- and modelling only coherent flows, as in the original study, or by retaining fingers-of-God features 
    and modelling them with a stochastic velocity dispersion component, as done by \citet{Hess_2013}.
    
    \item Forward modelling of RSDs at the dark-matter-field level.
    In this work, we followed the approach of mapping the matter field directly to redshift space in a fully differentiable Bayesian inference framework.
    We set the observer at the center of the cubical volume and computed the RSD effects accordingly along lines of sight. See \citet{Bos_2019} for a first implementation of such a method and calculation details.
\end{enumerate}

We note that the modelling of RSDs for biased tracers involving non-linear transformations -- such as cosmic voids and Lyman-$\alpha$ forests -- requires the inclusion of velocity bias. For the Lyman-$\alpha$ forest, RSD affects the optical depth, $\tau$, but the observable is the transmitted flux $F=\exp(-\tau)$, introducing a non-trivial mapping. Accurate modelling in these cases must account for velocity-bias effects (\citet[][]{McDonald_2000,Seljak_2012,Sinigaglia_2022,Sinigaglia_2024}). Similarly, cosmic voids are identified from the galaxy distribution in redshift space, which effectively applies a non-linear and non-local transformation (\citet[][]{Chuang_2017}). { For the time being, we restrict the implementation tests within the \wonj code to unbiased peculiar motions ($b_v=1$); i.e. the dark-matter-density contrast in redshift space is given by
\begin{equation}
  \boxed{  \delta(\vect{s})=\delta\left(\vect{r}+b_v\frac{\vect{v}\cdot\vect{r}}{aHr^2}\vect{r}\right)}  \,,
\end{equation}
with $\vect{s}$ and  $\vect{r}$ being the redshift- and real-space coordinates, respectively; $\vect{v}$ the peculiar velocity; $a$ the scale factor; and $H$ the Hubble constant for a given Hubble distance.
}

\section{Validation of field-level inference with fuzzy cosmic-web bias}
\label{sec:results}

To validate the \wonj code, we considered three numerical tests with different resolutions and bias models: \tone, \ttwo, and \tthree (see Table~\ref{tab:cases}). Results of \ttwo and \tthree are presented in Appendix~\ref{app:results}. 
The numerical tests consist of the following steps.
\begin{enumerate}
    \item Generate a ground-truth number-count catalogue of biased tracers  
    from an initial Gaussian density field with the same forward model employed during inference.
    All runs are performed to redshift $z=0$ with a particular random seed and specific bias parameters as listed in Tables \ref{tab:values} and \ref{tab:values3}.
    \item Perform field-level inference of the posterior describing the white-noise and bias parameters for \tone and \ttwo, applying the \wonj code.
    \item Perform quality assessment of the reconstructions by evaluating the recovery of both the initial and final conditions using two-point and three-point statistical analyses. 
\end{enumerate}

{
Regarding the hyperparameters of the bias parameter prior of Eq.~\ref{eq:lognormal}, the power-law ($\alpha$) and negative-binomial ($\beta$) hyperparameters are \(\mu_\alpha = 0.27, \sigma_\alpha = 0.30, \mu_\beta = 2.12, and \sigma_\beta = 0.20\).}

\begin{table}[h]
    \centering
    \caption{\label{tab:cases} Setting the different numerical cases.}
    \begin{tabular}{lccc}
        \toprule
                 & \tone & \ttwo & \tthree \\
        \midrule
        $\Delta L$ [\mpch] & 10  & 5   & 8          \\
        $L$ [\mpch] & 1280  & 640   & 1024        \\
        {$k_{\rm nyq}$} [\hmpc] & 0.31  & 0.63   & 0.39        \\
        {$k^{\rm non-iso}_{\rm nyq}$} [\hmpc] & 0.54  & 1.10   & 0.68        \\
        non-local bias  & $\Phi$-web$_j$  & $\Phi$-web$_j$   & $\Phi\delta$-web$_{jk}$        \\
        local bias  & \{$\alpha_j$, $\beta_j$\}  & \{$\alpha_j$, $\beta_j$\}  & $\{\alpha_{jk}$, $\beta_{jk}$, $\epsilon_{jk}$, $\rho_{jk}$\}     \\       
        \bottomrule
    \end{tabular}
    \tablefoot{In all cases, we used a mesh of $128^3$ voxels with cubical volumes of distinct side lengths, $L,$ and resolutions, ${\Delta}L=L/N$. We consider the Nyquist frequency $k_{\rm nyq}=\pi/\Delta L$ and the{ non-isotropic}  Nyquist frequency {$k^{\rm non-iso}_{\rm nyq}=\sqrt{3} \,k_{\rm nyq}$} of a cubical Fourier-space mesh. The output redshift is $z=0$. There are 16 combinations of $\Phi$- and $\delta$-web regions: $j,k\in[1,2,3,4]$.}
\end{table}

\begin{table}[]
    \centering
    \caption{\label{tab:values} Values of bias parameters by cosmic-web region of the reference for the \tone and \ttwo cases (8 bias parameters).}
    \begin{tabular}{l@{\hskip 10pt}lcccc}
        \toprule
        Test  &       & Voids & Sheets & Filaments & Knots \\
        \midrule
        \multirow{2}{*}{\centering \tone} & $\alpha$ & 1.10  & 1.03   & 1.15       & 1.23 \\
                                          & $\beta$  & 10.10  & 13.70   & 12.10       & 14.40 \\
        \midrule
        \multirow{2}{*}{\centering \ttwo} & $\alpha$ & 1.05  & 1.11   & 1.23       & 1.30 \\
                                          & $\beta$  & 7.10  & 8.70   & 8.10       & 9.40 \\
        \bottomrule
    \end{tabular}
\end{table}

\begin{table}[]
    \centering
    \caption{\label{tab:values3} Values of bias parameters by combined cosmic-web regions for the \tthree case (64 bias parameters).}
    \begin{tabular}{llcccc}
        \toprule
        $\Phi$-web & $\delta$-web & $\alpha$ & $\beta$ & $\epsilon$ & $\rho$ \\
        \midrule
        \multirow{4}{*}{Voids} 
            & Voids     & 1.29 & 5.80 & 1.02 & 0.22 \\
            & Sheets    & 1.05 & 14.30 & 1.06 & 0.26 \\
            & Filaments & 1.14 & 11.87 & 0.97 & 0.16 \\
            & Knots     & 1.34 & 10.60 & 1.00 & 0.20 \\
        \midrule
        \multirow{4}{*}{Sheets} 
            & Voids     & 1.04 & 10.74 & 1.00 & 0.20 \\
            & Sheets    & 1.06 & 11.57 & 0.97 & 0.17 \\
            & Filaments & 1.11 & 12.59 & 0.98 & 0.18 \\
            & Knots     & 1.05 & 13.67 & 1.06 & 0.26 \\
        \midrule
        \multirow{4}{*}{Filaments} 
            & Voids     & 0.94 & 5.71 & 1.02 & 0.22 \\
            & Sheets    & 1.13 & 13.42 & 0.93 & 0.13 \\
            & Filaments & 1.27 & 7.75 & 0.96 & 0.16 \\
            & Knots     & 1.20 & 11.21 & 1.01 & 0.21 \\
        \midrule
        \multirow{4}{*}{Knots} 
            & Voids     & 1.18 & 8.36 & 0.99 & 0.19 \\
            & Sheets    & 1.01 & 11.24 & 0.96 & 0.16 \\
            & Filaments & 1.07 & 10.99 & 0.94 & 0.14 \\
            & Knots     & 1.17 & 9.71 & 1.08 & 0.28 \\
        \bottomrule
    \end{tabular}
    \tablefoot{We show sub-divisions of $\delta$-web types within each $\Phi$-web classification (see Appendix \ref{app:types} and \citealt{Coloma_2024}).}
\end{table}

\subsection{Motivation of the numerical tests}

The numerical tests were chosen to demonstrate a number of scientific relevant cases. We list these below. 
\begin{itemize}
    \item \tone and \ttwo: The non-local-bias description used here is based on the $\Phi$-web classification. 
    \citet{Balaguera_2018} demonstrated that this model can accurately reproduce key statistical properties of the halo distribution, including one-point PDFs, two-point power spectra at percent-level accuracy, and bispectra with reasonable agreement, generally within 15\% (see also \citealt{Balaguera_2020}, \citealt{Pellejero_2020}, and \citealt{Balaguera_2023}, who confirm these results for different halo resolutions). These studies confirm the  $\Phi$ web as a physically meaningful and effective framework for modelling non-local bias within a reasonable degree of accuracy through non-parametric approaches. However, for field-level inference applications -- such as the one developed here -- a parametric bias model is required to enable efficient sampling and marginalisation within a Bayesian framework.
    The same non-local-bias treatment incorporating an equivalent non-linear-bias prescription to the one considered here (see Eq.~\ref{eq:local_bias}) was employed by \citet{Sinigaglia_2024,Sinigaglia_2024b} to model the Lyman-$\alpha$ forest. Their work demonstrated excellent agreement with high-resolution hydrodynamical simulations, achieving power-spectrum accuracy within 5\% up to $k\sim1$ \hmpc, along with consistent bispectrum predictions. In both cases, i.e. haloes and Lyman-$\alpha$ forests, the underlying structure-formation model was ALPT. For both the \tone and \ttwo cases, we jointly sampled the bias parameters starting with an initial guess randomly sampled from the prior distribution (see Eq.~\ref{eq:lognormal}).
    
    \item \tone focuses on a resolution of 10 \mpch, corresponding to a Nyquist frequency of $k\simeq0.3$ \hmpc. This resolution surpasses the typical scale used in most cosmological analyses, which are commonly limited to  $k<0.2$ \hmpc \citep[see e.g.][]{Ivanov_2025}. Also, the ideal resolution at which BAO reconstruction techniques are applied is slightly lower, i.e.15 \mpch\    \citep[see e.g.][]{Paillas_2025}.
    \item \ttwo focuses on a resolution of 5 \mpch{} with a Nyquist frequency of $k\simeq0.6$ \hmpc. This resolution has been reported to be high enough to produce accurate galaxy catalogues combined with a sub-grid model based on ALPT \citep[see][]{Forero_2024}. 
    
    \item \tthree: This case study incorporates the full HICOBIAN non-local bias model, which has been shown to significantly enhance accuracy in both power spectra and bispectra, achieving full compatibility with $N$-body-based catalogues across a wide range of scales. While \citet{Coloma_2024} demonstrated that this model remains valid at mesh resolutions below 4 \mpch, in this study we restricted our analysis to a coarser resolution of 8 \mpch. This choice is motivated by the fact that the corresponding Nyquist frequency, $k\simeq0.4$ \hmpc, marks the scale beyond which cosmological information is effectively saturated \citep[][]{Hahn_2021}. For computational reasons, we kept the bias parameters fixed in this case, assuming they are known. This assumption is reasonable if the parameters can be extracted from high-fidelity reference catalogues, as was done for the PATCHY BOSS mocks, where the MULTIDARK reference catalogue was designed to reproduce the observed galaxy distribution with halo abundance matching \citep[][]{Torres_2016}. Even more sophisticated reference catalogues are now being developed to reproduce the clustering statistics of the observed DESI catalogues \citep[][]{DESI_2025}, from which the parameters of the HICOBIAN model can be accurately extracted \citep[][]{Favole_2025}.
\end{itemize}
 We considered $\rho'=0$ and $\epsilon'=1$ for simplicity, interpreting this by continuity for $\delta>-1$, so the high-density damping factor, $\exp \left[-
      \left(
      \frac{1 + \delta_i}{\rho'_\text{}}
      \right)^{ -\epsilon'_{\text{}}}
      \right]$, was set to one in all three test cases.

\subsection{Numerical results of \tone}

The first numerical test was performed on a grid with \(128^3\) voxels within a comoving volume of 1280~\mpch\ per side, corresponding to a spatial resolution of 10~\mpch. 
We used a $\Phi$-web-dependent bias modelled with a power-law and negative-binomial likelihood for each region, for a total of eight bias parameters. 
Using this model and a fixed random seed, we constructed a mock catalogue of object number counts, $\vect{n}^*$, in redshift space, which we define as the ground truth. 
The priors of the power-law ($\alpha$) and negative-binomial ($\beta$) bias parameters were set to be log-normally distributed  (see Eq~\ref{eq:lognormal}) with fixed hyperparameters: \(\mu_\alpha = 0.27, \sigma_\alpha = 0.30, \mu_\beta = 2.12, \sigma_\beta = 0.20\).

The HMC chain was initialised with an state $(\vect{\nu}^{(0)}, \vect{b}^{(0)})$, with $\vect{\nu}^{(0)}$ being the white-noise representation of the density contrast coming from Wiener-filtering, $\vect{n}^*$, and $\vect{b}^{(0)}$ being bias parameters randomly sampled from their priors.  During burn-in, chain convergence was achieved after approximately 250 samples. The adapted step size was $ \sim \! 10^{-3}$, and the NUTS trajectory length adaptation saturated at 1024 leapfrog steps (corresponding to a maximum tree depth of 10). 

\begin{figure}[h!]
\centering
\includegraphics[width=1\columnwidth]{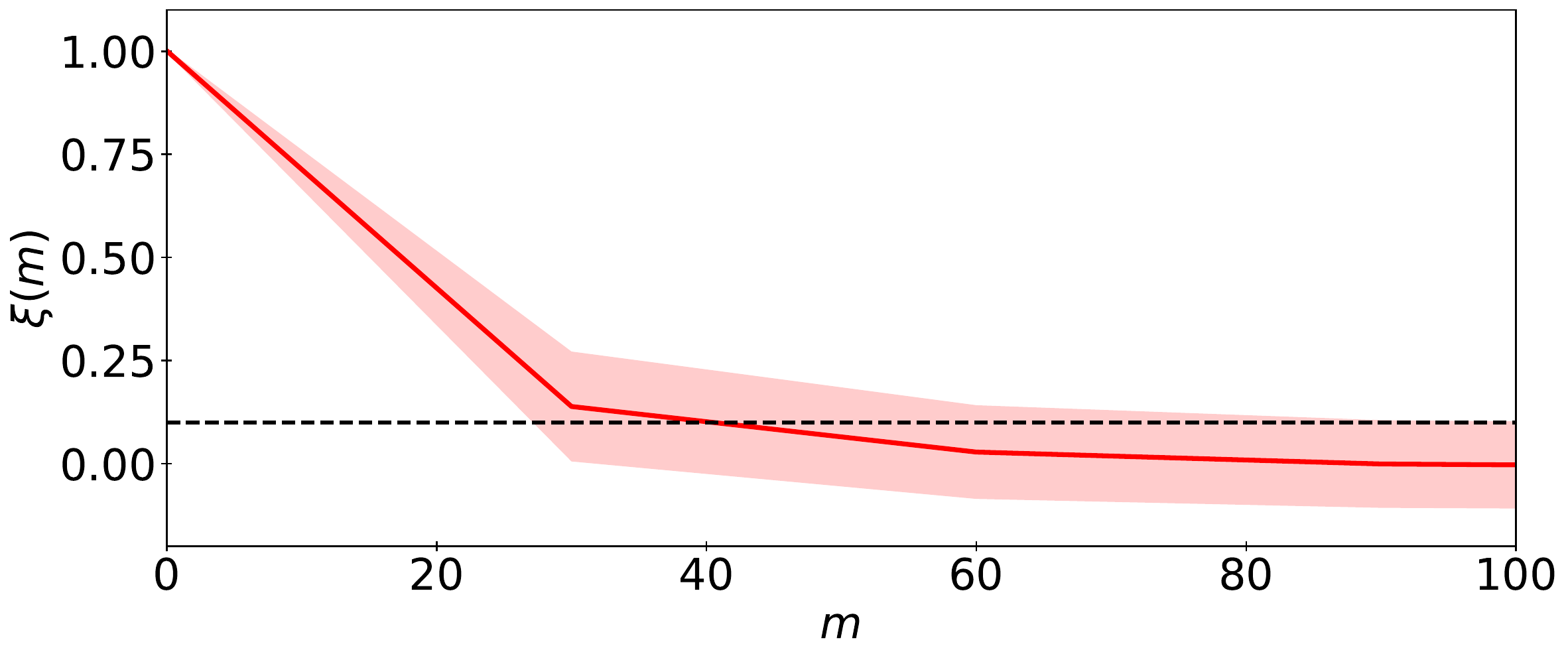}
    \caption{\tone: Mean auto-correlation $\xi(n)$ as a function of lag, $n,$ for a chain of length of $M\approx1000$, evaluated over $10^4$ randomly selected voxels. The  red curve shows the ensemble average, $\langle\xi(n)\rangle,$ across parameters, with the shaded band indicating the $\pm1\sigma$ dispersion. The horizontal dashed line marks the threshold, $\xi=0.1,$ used to assess mixing efficiency.}
    \label{fig:corr_length}
\end{figure}

After convergence, we assessed the correlation length in the white-noise  samples, based on their autocorrelation at lag {\(m\)}: 
\begin{equation}
    \xi_i(m) =
    \frac{1}{\sigma^2_i {\left(M-m\right)}}
    \sum_{j=0}^{M-m}
    \left(\nu_i^{(j)} - \bar{\nu}_i\right)
    \left(\nu_i^{(j+m)} - \bar{\nu}_i\right)
    \ , 
\end{equation}
where $\nu_i^{(j)}$ is the $i$th voxel of the $j$th sample, \(M\) the total number of samples of a given chain run,
\(\bar{\nu}_i\) the within-chain mean, and \(\sigma_i^2\) the within-chain variance. 
For a chain of \(M = 1000\) samples, we computed \(\xi_i(m)\) for lags {\(m = 0\)} to \(100\) across \(10^4\) randomly selected voxels.
Figure~\ref{fig:corr_length} shows the mean auto-correlation over this subset.

We defined the correlation length as the smallest lag, \(n,\) such that \(\langle \xi(n) \rangle \leq 0.1\), yielding an estimate of $\sim\!40$ samples. Given this auto-correlation, we drew 15000 post-burn-in states, providing about 500 effectively independent posterior samples for the subsequent statistical analysis.

\begin{figure}[h!]
\centering
\includegraphics[width=1\columnwidth]{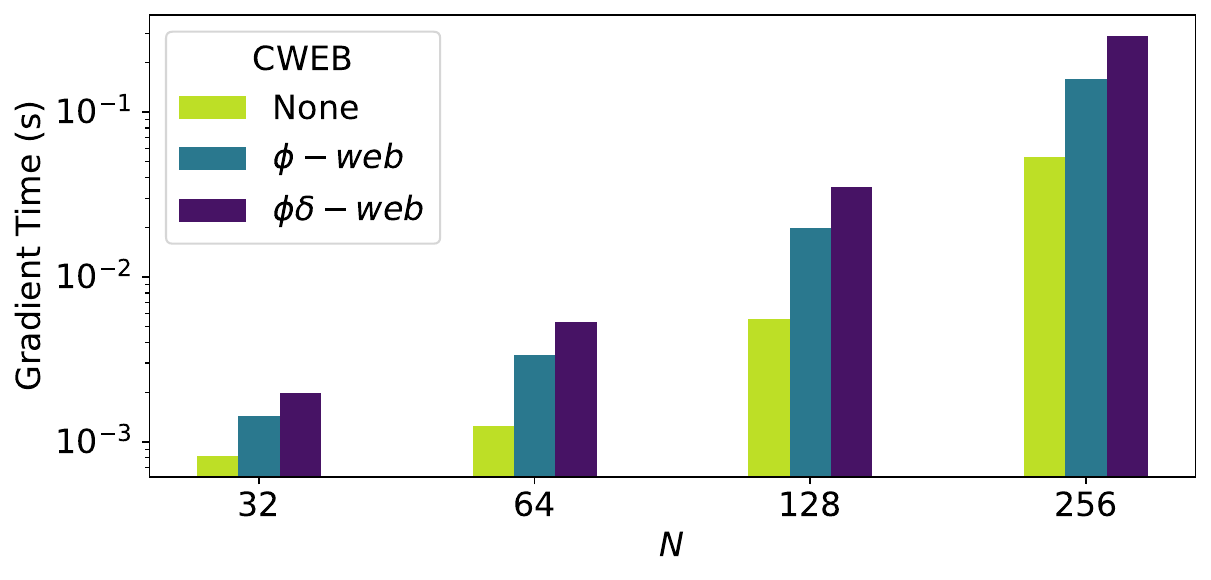}
\caption{Computing times for single-gradient evaluations of the forward model employing ALPT evolution and the HICOBIAN bias model for different mesh sizes, run on a single NVIDIA A100-SXM4 GPU equipped with 40 GB of on-board HBM2 memory.{ For reference, the full burn-in and sampling procedure for the $128^3$ case in this study required approximately 100 GPU hours.} }       
\label{fig:stress_test}
\end{figure}

\begin{figure}[h!]
\centering \includegraphics[width=1\columnwidth]{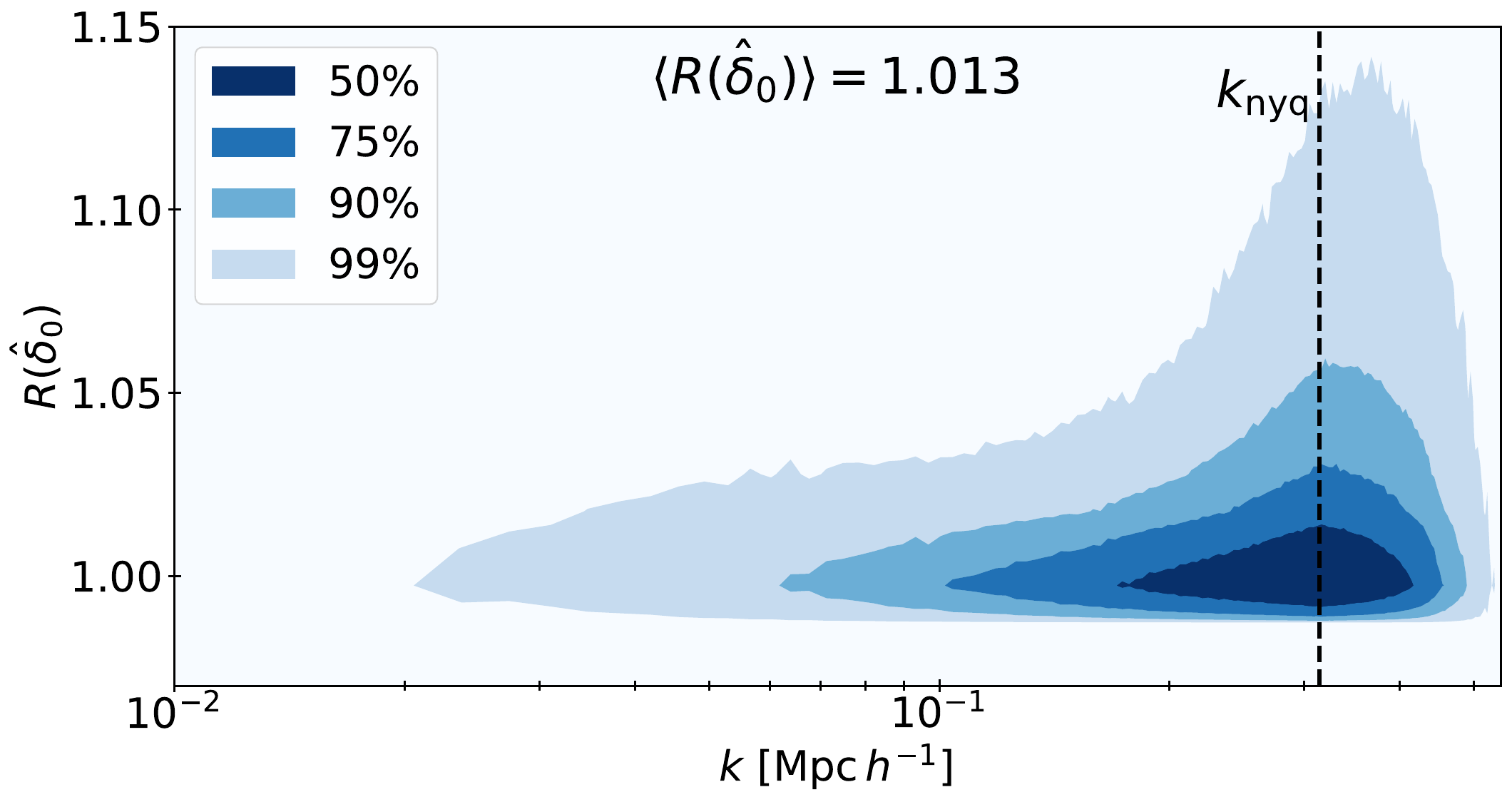}\caption{Gelman--Rubin convergence diagnostic computed for three independently initialised chains of 250 samples  after convergence. This statistic is calculated for all cells in Fourier space, and we show their distribution with scale, $k$. Values close to one indicate that the drawn samples have converged to the same distribution.
}
\label{fig:gr_test}
\end{figure}

We assessed the computing times of single-gradient calculations across different mesh sizes and bias configurations finding that they are below one second (see Fig.~\ref{fig:stress_test}). 
In addition, we performed the Gelman--Rubin convergence diagnostic, which confirms that the drawn samples{ have converged to the same target distribution} (see Fig.~\ref{fig:gr_test}).

\begin{figure*}[p]
\centering
\includegraphics[width=0.8\textwidth]{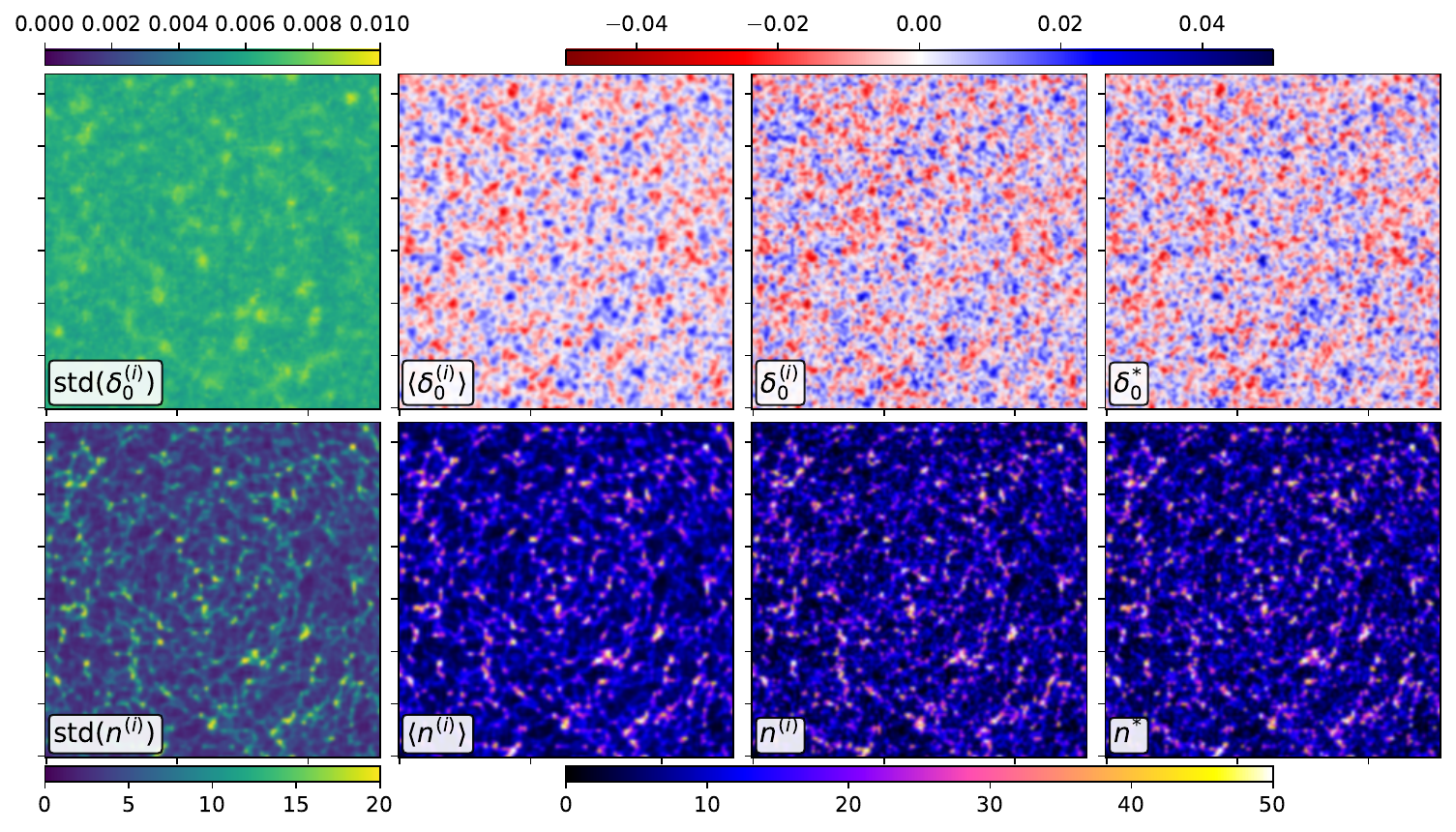}
    \caption{\label{fig:maps}\tone: 
        Spatial summary of 500 independent reconstructions ($N=128$). Slices of $\Delta L=10\,h^{-1}\mathrm{Mpc}$ (one voxel width). Top row: Reconstructed initial density contrast. Bottom row: Reconstructed tracer field. Columns from left to right: (1) Pixel-wise standard deviation across the 500 independent reconstructions; (2) mean reconstructed field; (3) one representative reconstruction sample; and (4) true reference field.
    }
\end{figure*}

As a first step, we conducted a visual assessment of the true and reconstructed maps of the initial and final density fields (see Fig.~\ref{fig:maps}). The reconstructed samples exhibit a high degree of visual similarity to the ground truth for both the initial and final conditions. 
On a more quantitative level, we find that the standard deviations of the reconstructed fields display structured patterns at amplitudes approximately an order of magnitude lower than the true density field at the initial conditions. In contrast, the final conditions exhibit higher variance, reflecting the increased non-linearity and complexity of the evolved structures.

\begin{figure*}[p]
\centering
\includegraphics[width=1\textwidth]{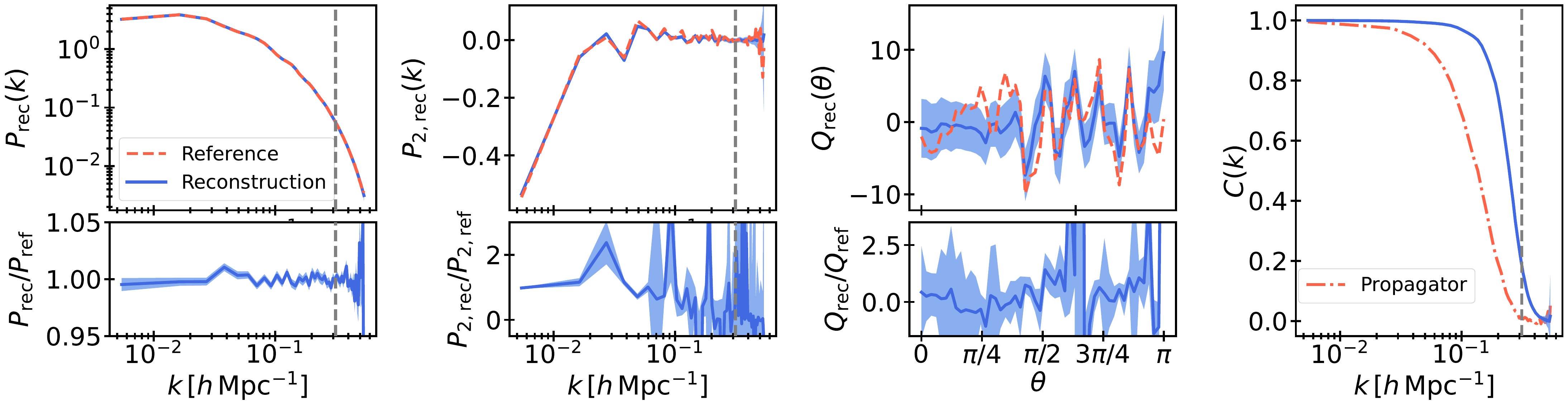}
    \caption{\label{fig:din_stats}\tone:  Comparison of summary statistics of the initial density contrast field for the mock reference (dashed red line) and 500 independent reconstructions (solid blue line).  Columns from left to right: (1) Monopole power spectrum; (2) quadrupole; (3) reduced bispectrum as a function of triangle opening angle, $\theta$; and (4) cross-correlation coefficient $C(k)$ between the reference and reconstruction (solid blue line) and the propagator (dashed red line), defined as the cross-correlation between the mock tracer density contrast and the initial density reference. In columns (1)--(3), the lower sub-panels display the ratios with respect to the reference. Shaded bands denote the $1\sigma$ intervals from the posterior samples. Vertical dashed lines indicate the isotropic Nyquist frequency, $k_\mathrm{nyq}/\sqrt{3}$.
        }
\end{figure*}

\begin{figure*}[p]
\centering
\includegraphics[width=1\textwidth]{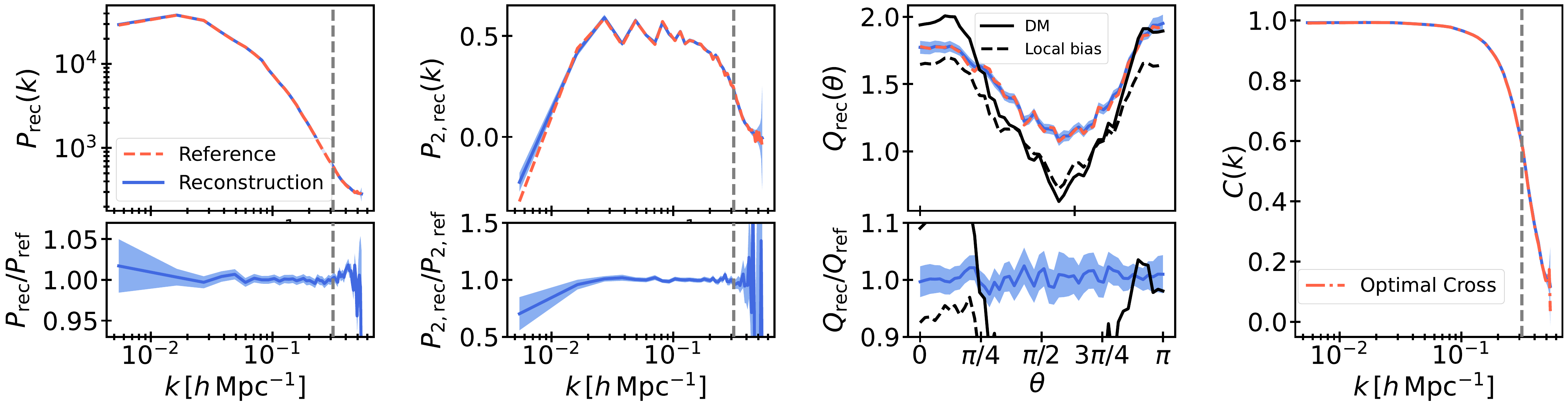}
    \caption{\label{fig:ntr_stats}\tone:  Comparison of summary statistics of the mock reference and 500 independent reconstructions evolved through the forward model. Panel definitions are identical to those in Fig.~\ref{fig:din_stats}, except that in the rightmost panel the propagator is replaced by the maximal tracer field cross-correlation and that we show the bispectrum of the dark-matter field and a tracer field using a local bias (without cosmic web) with bias parameters set to those of knots.     
    }
\end{figure*}

To quantitatively assess the quality of the reconstruction, we computed both two- and three-point statistics in Fourier space for the initial (Fig.~\ref{fig:din_stats}) and final conditions (Fig.~\ref{fig:ntr_stats}). Specifically, we evaluated the monopole and quadrupole moments of the power spectrum, along with the reduced bispectrum for a configuration particularly sensitive to non-linear effects ($k_1=0.1$ and $k_2=0.2$ \hmpc). In addition, we computed the propagator, { defined as the cross-correlation between the mock tracer density contrast and the initial density reference.}
The motivation for investigating the quadrupole and bispectrum of a the reconstructed Gaussian field was to assess how well the samples recover deviations introduced by cosmic variance in these statistics. 

In Figs.~\ref{fig:din_stats} and \ref{fig:ntr_stats}, it is remarkable to observe that the monopoles, quadrupoles, and bispectra are accurately recovered not only in the final conditions, but also in the initial conditions, even reproducing the specific features inherent to the particular cosmic variance realisation. Although local-bias models fail to reproduce three-point statistics without short- and long-range non-local terms \citep[][]{Coloma_2024}, we show that only neglecting long-range non-local bias also leads to significant differences in the bispectrum (see the third panel in Fig.~\ref{fig:ntr_stats}). 

The propagator ensures a significant gain in information, demonstrating the effectiveness of the reconstruction. To evaluate how close the result is to an ideal scenario, we computed the optimal cross at final conditions by generating a synthetic-tracer number-count field from the evolved ground-truth dark-matter field using the exact same bias parameters and stochastic prescription as in the model, but a different seed. The cross-correlation between this ideal tracer field and the ground truth serves as a benchmark, represented by the dashed-dotted red line in the figure. This allowed us to assess whether the reconstruction approaches the theoretical limit given the assumed bias model. As we see in Fig.~\ref{fig:ntr_stats}, the red and blue lines of the right panel are on top of each other, indicating that the optimal cross-correlation limited by shot noise has been reached.  

\begin{figure*}[h!]
\centering
\includegraphics[width=1\textwidth]{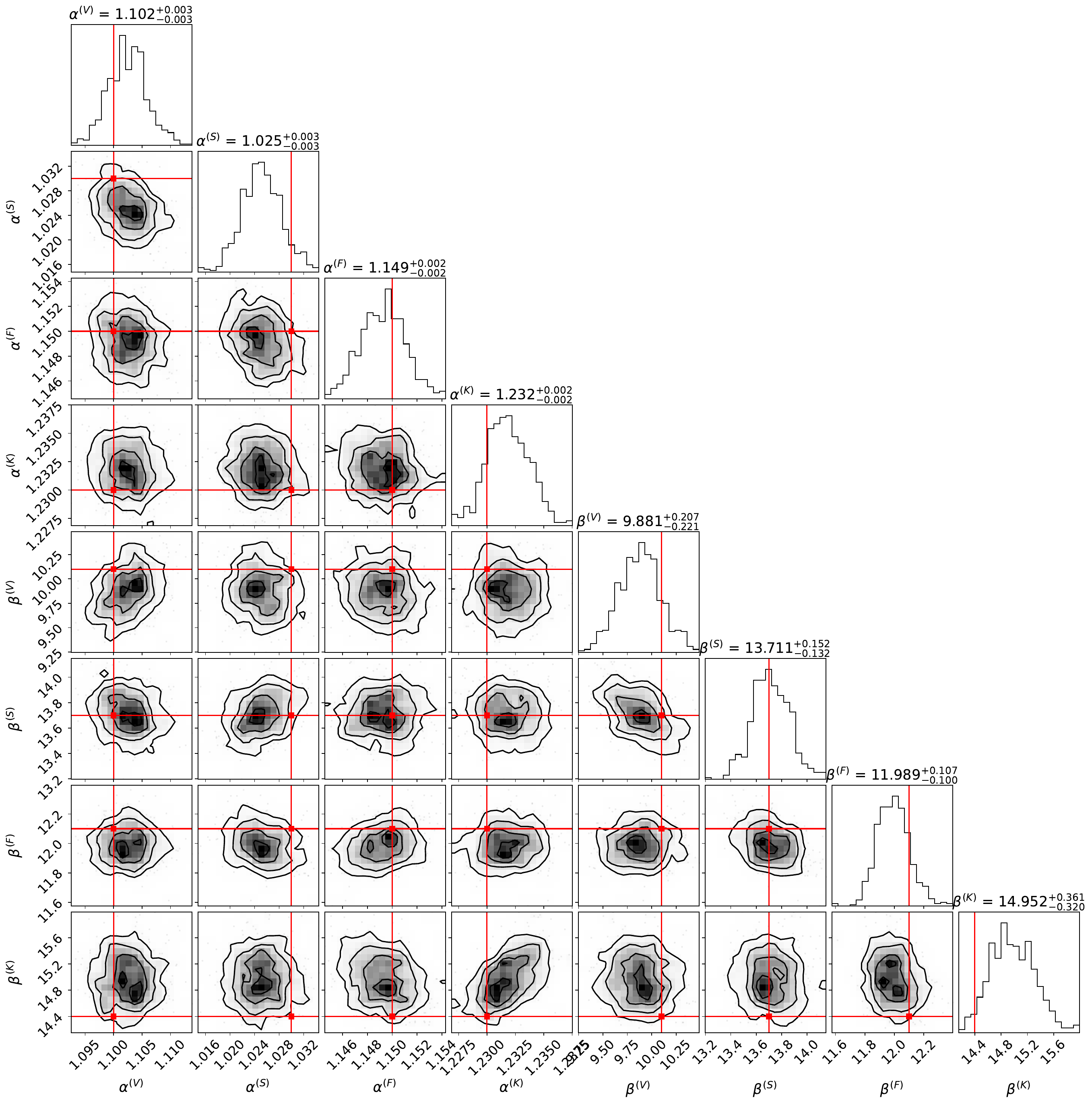}
    \caption{\tone: Posterior distributions for the 500 independent samples of the $\alpha$  and $\beta$ bias parameters in each cosmic-web region of the $\Phi$ web. Red dots and lines show the values of the parameters for the ground-truth catalogue. The posteriors are smoothed with a 2D Gaussian filter with standard deviation equal to 0.75 bin widths. \ttwo yields equivalent results.}
\label{fig:corner_plot_test1}
\end{figure*}

In Fig.~\ref{fig:corner_plot_test1}, we show a corner plot of the posterior distributions of the eight parameters of the bias model compared to their values in the reference catalogue. The power-law parameters, $\alpha,$ were recovered with sub-percent accuracy, while the negative binomial $\beta$-parameter was recovered with precision of a few percent. The absence of significant correlations between $\alpha$ bias parameters in distinct cosmic-web regions supports the validity and stability of our inference framework.

Overall, the consistency of all diagnostics -- maps, summary statistics, and bias-parameter posteriors -- demonstrates that the complete field-level inference pipeline reliably recovers both the underlying density field and the cosmic-web-dependent bias parameters at the targeted resolution. The numerical studies \ttwo and \tthree further confirm the validation of the \wonj code (details 
can be found in Appendix~\ref{app:results}). However, it is remarkable that we did not recover the original bias parameters
when we allowed them to vary freely in case \tthree{}.

{  }

\section{Conclusions}

In this work we developed and validated a novel Bayesian field-level inference framework that incorporates a physically motivated, differentiable, and non-local bias model -- HICOBIAN -- based on the hierarchical cosmic-web. This model allows for smooth transitions between different cosmic environments via a fuzzy classification approach, making it well suited for machine-learning integration and GPU acceleration.

We demonstrated that, within a self-specified setting where inference is performed on data generated by the exact same forward model, our approach accurately reconstructs the initial density field, achieving high fidelity even in the presence of redshift-space distortions and complex, non-linear, non-local, and stochastic bias. The reconstruction quality was evaluated using two- and three-point statistics, including the power-spectrum monopole and quadrupole, the reduced bispectrum, and the propagator, showing excellent agreement with the ground truth and reaching the optimal limit set by shot noise.

We further tested the robustness of the model under various conditions, including varying spatial resolutions and complex bias scenarios involving four to 16 cosmic-web regions, each characterised by distinct bias parameters. These included power-law bias, exponential threshold bias, and deviations from Poisson statistics, totalling eight to 64 bias parameters.  Across all settings, the method maintained its high accuracy, confirming the generality and adaptability of our approach. However, for the complex bias models, the original bias parameters were
not recovered when left to vary freely. This hints at the complexity of bias
modelling and the potential limitations of field-level inference approaches
that ignore intra-halo dynamics. As demonstrated in recent work \citep[e.g.][]{Favole_2025},
a realistic halo-to-galaxy connection based on high-resolution $N$-body simulations
requires a large number of parameters to capture processes such as halo exclusion,
satellite dynamics, quenching, and intra-halo flows.

{ }

In future work, we plan to investigate the performance of the \wonj code with high-fidelity synthetic reference catalogues and directly apply it to redshift survey data across various tracers, including { DESI-tracers, such as} bright galaxies, luminous red galaxies, emission-line galaxies, quasars, and Lyman-$\alpha$ forests. This will require the incorporation of light-cone evolution and selection effects and a more detailed analysis of redshift-space distortions  in the highly non-linear regime.{ Moreover, our results suggest that achieving full consistency may require a joint Bayesian inference framework that incorporates intra-halo dynamics informed by detailed $N$-body simulations to break the degeneracy in the bias parameters.}

By implementing this framework in the GPU-accelerated \wonj code using \jax, we enable efficient and scalable inference on large datasets. This work thus marks a significant step towards a unified, interpretable, and fully differentiable framework for cosmological data analysis, with direct applicability to tracers of the large-scale structure.

\begin{acknowledgements}
The authors thank the Spanish Ministry of Science and Innovation for funding the \texttt{Big Data of the Cosmic-Web} project (PID2020-120612GB-I00 / AEI / 10.13039/501100011033, PI:FSK), under which this work was conceived and carried out. We are also grateful to the Instituto de Astrofísica de Canarias (IAC) for its continued support through the \texttt{Cosmology with LSS Probes} project (PI:FSK).
The authors thank Jesús Sanz Serna and José María Coloma-Nadal for discussions. 
PR acknowledges support from the IAC Research Summer Grant 2024  for the project ``Bayesian Inference of the LSS with ALPT and HICOBIAN bias using JAX''. This grant enabled his contribution to this work prior to the start of his master’s thesis on the same topic, as reflected in this manuscript. He also thanks Gustavo Yepes, the Universidad Autónoma de Madrid (UAM), and the Department of Theoretical Physics for their hospitality and for providing access to computing facilities during his visit -- resources that were instrumental to the development of this work.
\end{acknowledgements}

\bibliographystyle{aa}
\bibliography{AA55817-25} 

\begin{thebibliography}{128}
\expandafter\ifx\csname natexlab\endcsname\relax\def\natexlab#1{#1}\fi

\bibitem[{{Amendola} {et~al.}(2016){Amendola}, {Appleby}, {Avgoustidis}, \& {others}}]{Euclid}
{Amendola}, L., {Appleby}, S., {Avgoustidis}, A., \& {others}. 2016 [\eprint[arXiv]{1606.00180}]

\bibitem[{Andrews {et~al.}(2023)Andrews, Jasche, Lavaux, \& Schmidt}]{andrewsBayesianFieldlevelInference2023}
Andrews, A., Jasche, J., Lavaux, G., \& Schmidt, F. 2023, MNRAS, 520, 5746

\bibitem[{{Angulo} \& {Hahn}(2022)}]{Angulo_2022}
{Angulo}, R.~E. \& {Hahn}, O. 2022, LRCA, 8, 1

\bibitem[{{Ata} {et~al.}(2021){Ata}, {Kitaura}, {Lee}, {Lemaux}, {Kashino}, {Cucciati}, {Hern{\'a}ndez-S{\'a}nchez}, \& {Le F{\`e}vre}}]{Ata_2021}
{Ata}, M., {Kitaura}, F.-S., {Lee}, K.-G., {et~al.} 2021, \mnras, 500, 3194

\bibitem[{{Ata} {et~al.}(2015){Ata}, {Kitaura}, \& {M{\"u}ller}}]{Ata_2015}
{Ata}, M., {Kitaura}, F.-S., \& {M{\"u}ller}, V. 2015, MNRAS, 446, 4250

\bibitem[{{Balaguera-Antol{\'\i}nez} {et~al.}(2023){Balaguera-Antol{\'\i}nez}, {Kitaura}, {Alam}, {Chuang}, {Yu}, {Favole}, {Sinigaglia}, {Zhao}, {Brooks}, {de la Macorra}, {Font-Ribera}, {Gontcho A Gontcho}, {Honscheid}, {Kehoe}, {Meisner}, {Miquel}, {Tarl{\'e}}, {Vargas-Maga{\~n}a}, \& {Zhou}}]{Balaguera_2023}
{Balaguera-Antol{\'\i}nez}, A., {Kitaura}, F.-S., {Alam}, S., {et~al.} 2023, \aap, 673, A130

\bibitem[{{Balaguera-Antol{\'\i}nez} {et~al.}(2020){Balaguera-Antol{\'\i}nez}, {Kitaura}, {Pellejero-Ib{\'a}{\~n}ez}, {Lippich}, {Zhao}, {S{\'a}nchez}, {Dalla Vecchia}, {Angulo}, \& {Crocce}}]{Balaguera_2020}
{Balaguera-Antol{\'\i}nez}, A., {Kitaura}, F.-S., {Pellejero-Ib{\'a}{\~n}ez}, M., {et~al.} 2020, MNRAS, 491, 2565

\bibitem[{Balaguera-Antolínez {et~al.}(2018)Balaguera-Antolínez, Kitaura, Pellejero-Ibáñez, Zhao, \& Abel}]{Balaguera_2018}
Balaguera-Antolínez, A., Kitaura, F.-S., Pellejero-Ibáñez, M., Zhao, C., \& Abel, T. 2018, MNRAS: Letters, 483, L58

\bibitem[{{Baldauf} {et~al.}(2013){Baldauf}, {Seljak}, {Smith}, {Hamaus}, \& {Desjacques}}]{Baldauf_2013}
{Baldauf}, T., {Seljak}, U., {Smith}, R.~E., {Hamaus}, N., \& {Desjacques}, V. 2013, \prd, 88, 083507

\bibitem[{{Bayer} {et~al.}(2023){Bayer}, {Seljak}, \& {Modi}}]{Bayer2023}
{Bayer}, A.~E., {Seljak}, U., \& {Modi}, C. 2023, arXiv e-prints, arXiv:2307.09504

\bibitem[{{Bernardeau}(1994)}]{Bernardeau_1994}
{Bernardeau}, F. 1994, \apj, 427, 51

\bibitem[{Bernardeau {et~al.}(2002)Bernardeau, Colombi, Gazta{\~{n}}aga, \& Scoccimarro}]{Bernardeau_2002}
Bernardeau, F., Colombi, S., Gazta{\~{n}}aga, E., \& Scoccimarro, R. 2002, Physics Reports, 367, 1

\bibitem[{{Bertschinger} \& {Dekel}(1989)}]{Bertschinger_1989}
{Bertschinger}, E. \& {Dekel}, A. 1989, \apjl, 336, L5

\bibitem[{{Bi} \& {Davidsen}(1997)}]{Bi_1997}
{Bi}, H. \& {Davidsen}, A.~F. 1997, \apj, 479, 523

\bibitem[{{Blanes} {et~al.}(2014){Blanes}, {Casas}, \& {Sanz-Serna}}]{Blanes_2014}
{Blanes}, S., {Casas}, F., \& {Sanz-Serna}, J.~M. 2014, SIAM Journal on Scientific Computing, 36, A1556

\bibitem[{{Blot} {et~al.}(2019){Blot}, {Crocce}, {Sefusatti}, {Lippich}, {S{\'a}nchez}, {Colavincenzo}, {Monaco}, {Alvarez}, {Agrawal}, {Avila}, {Balaguera-Antol{\'\i}nez}, {Bond}, {Codis}, {Dalla Vecchia}, {Dorta}, {Fosalba}, {Izard}, {Kitaura}, {Pellejero-Ibanez}, {Stein}, {Vakili}, \& {Yepes}}]{Blot_2019}
{Blot}, L., {Crocce}, M., {Sefusatti}, E., {et~al.} 2019, MNRAS, 485, 2806

\bibitem[{{Bond} {et~al.}(1996){Bond}, {Kofman}, \& {Pogosyan}}]{Bond_1996nat}
{Bond}, J.~R., {Kofman}, L., \& {Pogosyan}, D. 1996, \nat, 380, 603

\bibitem[{{Bos} {et~al.}(2019){Bos}, {Kitaura}, \& {van de Weygaert}}]{Bos_2019}
{Bos}, E.~G.~P., {Kitaura}, F.-S., \& {van de Weygaert}, R. 2019, MNRAS, 488, 2573

\bibitem[{{Bouchet} {et~al.}(1995){Bouchet}, {Colombi}, {Hivon}, \& {Juszkiewicz}}]{Bouchet_1995}
{Bouchet}, F.~R., {Colombi}, S., {Hivon}, E., \& {Juszkiewicz}, R. 1995, \aap, 296, 575

\bibitem[{Bradbury {et~al.}(2018)Bradbury, Frostig, Hawkins, Johnson, Leary, Maclaurin, Necula, Paszke, Vander{P}las, Wanderman-{M}ilne, \& Zhang}]{jax2018github}
Bradbury, J., Frostig, R., Hawkins, P., {et~al.} 2018, http://github.com/jax-ml/jax

\bibitem[{{Buchert}(1994)}]{Buchert_1994}
{Buchert}, T. 1994, MNRAS, 267, 811

\bibitem[{Casas-Miranda {et~al.}(2002)Casas-Miranda, Mo, Sheth, \& Boerner}]{Casas_Miranda_2002}
Casas-Miranda, R., Mo, H.~J., Sheth, R.~K., \& Boerner, G. 2002, MNRAS, 333, 730–738

\bibitem[{{Catelan}(1995)}]{Catelan_1995}
{Catelan}, P. 1995, MNRAS, 276, 115

\bibitem[{{Cen} \& {Ostriker}(1993)}]{Cen_1993}
{Cen}, R. \& {Ostriker}, J.~P. 1993, \apj, 417, 415

\bibitem[{{Chan} {et~al.}(2012){Chan}, {Scoccimarro}, \& {Sheth}}]{Chan_2012}
{Chan}, K.~C., {Scoccimarro}, R., \& {Sheth}, R.~K. 2012, \prd, 85, 083509

\bibitem[{{Chuang} {et~al.}(2017){Chuang}, {Kitaura}, {Liang}, {Font-Ribera}, {Zhao}, {McDonald}, \& {Tao}}]{Chuang_2017}
{Chuang}, C.-H., {Kitaura}, F.-S., {Liang}, Y., {et~al.} 2017, \prd, 95, 063528

\bibitem[{{Chuang} {et~al.}(2019){Chuang}, {Yepes}, {Kitaura}, {Pellejero-Ibanez}, {Rodr{\'\i}guez-Torres}, {Feng}, {Metcalf}, {Wechsler}, {Zhao}, {To}, {Alam}, {Banerjee}, {DeRose}, {Giocoli}, {Knebe}, \& {Reyes}}]{Chuang_2019}
{Chuang}, C.-H., {Yepes}, G., {Kitaura}, F.-S., {et~al.} 2019, MNRAS, 487, 48

\bibitem[{{Coles} \& {Jones}(1991)}]{Coles_1991}
{Coles}, P. \& {Jones}, B. 1991, MNRAS, 248, 1

\bibitem[{{Coloma-Nadal} {et~al.}(2024){Coloma-Nadal}, {Kitaura}, {Garc{\'\i}a-Farieta}, {Sinigaglia}, {Favole}, \& {Forero S{\'a}nchez}}]{Coloma_2024}
{Coloma-Nadal}, J.~M., {Kitaura}, F.-S., {Garc{\'\i}a-Farieta}, J.~E., {et~al.} 2024, JCAP, 2024, 083

\bibitem[{{Concei{\c{c}}{\~a}o} {et~al.}(2024){Concei{\c{c}}{\~a}o}, {Krone-Martins}, {da Silva}, \& {Molin{\'e}}}]{Conceiccao_2024}
{Concei{\c{c}}{\~a}o}, M., {Krone-Martins}, A., {da Silva}, A., \& {Molin{\'e}}, {\'A}. 2024, \aap, 681, A123

\bibitem[{{Daly} \& {Gaunt}(2016)}]{Daly2015}
{Daly}, F. \& {Gaunt}, R.~E. 2016, ALEA: Latin American Journal of Probability and Mathematical Statistics, 13, 635

\bibitem[{{Dekel} \& {Lahav}(1999)}]{Dekel_1999}
{Dekel}, A. \& {Lahav}, O. 1999, ApJ, 520, 24

\bibitem[{{DESI Collaboration} {et~al.}(2025{\natexlab{a}}){DESI Collaboration}, {Abdul-Karim}, {Adame}, {Aguado}, {Aguilar}, {Ahlen}, {Alam}, {Aldering}, {Alexander}, {Alfarsy}, {Allen}, \& {Allende Prieto}}]{DESI_2025}
{DESI Collaboration}, {Abdul-Karim}, M., {Adame}, A.~G., {et~al.} 2025{\natexlab{a}}, arXiv:2503.14745

\bibitem[{{DESI Collaboration} {et~al.}(2025{\natexlab{b}}){DESI Collaboration}, {Abdul-Karim}, {Adame}, {Aguado}, {Aguilar}, {Ahlen}, {Alam}, {Aldering}, {Alexander}, {Alfarsy}, {Allen}, {Allende Prieto}, {Alves}, {Anand}, {Andrade}, {Armengaud}, {Avila}, {Aviles}, {Awan}, {Bailey}, {Baleato Lizancos}, {Ballester}, {Bault}, {Bautista}, {BenZvi}, {Beraldo e Silva}, {Bermejo-Climent}, {Beutler}, {Bianchi}, {Blake}, {Blum}, {Bolton}, {Bonici}, {Brieden}, {Brodzeller}, {Brooks}, {Buckley-Geer}, {Burtin}, {Canning}, {Carnero Rosell}, {Carr}, {Carrilho}, {Casas}, {Castander}, {Cereskaite}, {Cervantes-Cota}, {Chaussidon}, {Chaves-Montero}, {Chen}, {Chen}, {Claybaugh}, {Cole}, {Cooper}, {Cousinou}, {Cuceu}, {Davis}, {Dawson}, {de Belsunce}, {de la Cruz}, {de la Macorra}, {de Mattia}, {Deiosso}, {Della Costa}, {Demina}, {Demirbozan}, {DeRose}, {Dey}, {Dey}, {Ding}, {Ding}, {Doel}, {Douglass}, {Dowicz}, {Ebina}, {Edelstein}, {Eisenstein}, {Elbers}, {Emas}, {Escoffier}, {Fagrelius}, {Fan}, {Fanning}, {Fawcett},
  {Fern\'andez-Garc\'ia}, {Ferraro}, {Findlay}, {Font-Ribera}, {Forero-Romero}, {Forero-S\'anchez}, {Frenk}, {G\''ansicke}, {Galbany}, {Garc\'ia-Bellido}, {Garcia-Quintero}, {Garrison}, {Gazta\\raisebox{-0.5ex}\~naga}, {Gil-Mar\'in}, {Gnedin}, {Gontcho}, {Gonzalez-Morales}, {Gonzalez-Perez}, {Gordon}, {Graur}, {Green}, {Gruen}, {Gsponer}, {Guandalin}, {Gutierrez}, {Guy}, {Hahn}, {Han}, {Han}, {He}, {Herrera-Alcantar}, {Honscheid}, {Hou}, {Howlett}, {Huterer}, {Ir\v\{s\}i\v\{c\}}, {Ishak}, {Jacques}, {Jimenez}, {Jing}, {Joachimi}, {Joudaki}, {Joyce}, {Jullo}, {Juneau}, {Kara\c\{c\}ayl\{\i\}}, {Karim}, {Kehoe}, {Kent}, {Khederlarian}, {Kirkby}, {Kisner}, {Kitaura}, {Kizhuprakkat}, {Kong}, {Koposov}, {Kremin}, {Krolewski}, {Lahav}, {Lai}, {Lamman}, {Lan}, {Landriau}, {Lang}, {Lange}, {Lasker}, {Le Goff}, {Le Guillou}, {Leauthaud}, {Levi}, {Li}, {Li}, {Lodha}, {Lokken}, {Luo}, {Magneville}, {Manera}, {Manser}, {Margala}, {Martini}, {Maus}, {McCullough}, {McDonald}, {Medina}, {Medina-Varela}, {Meisner},
  {Mena-Fern\'andez}, {Menegas}, {Mezcua}, {Miquel}, {Montero-Camacho}, {Moon}, {Moustakas}, {Mu\\raisebox{-0.5ex}\~noz-Guti\'errez}, {Mu\\raisebox{-0.5ex}\~noz-Santos}, {Myers}, {Myles}, {Nadathur}, {Najita}, {Napolitano}, {Newman}, {Nikakhtar}, {Nikutta}, {Niz}, {Noriega}, {Padmanabhan}, {Paillas}, {Palanque-Delabrouille}, {Palmese}, {Pan}, {Pan}, {Parkinson}, {Peacock}, {Percival}, {P\'erez-Fern\'andez}, {P\'erez-R\`afols}, \& {Peterson}}]{DESI_DR1}
{DESI Collaboration}, {Abdul-Karim}, M., {Adame}, A.~G., {et~al.} 2025{\natexlab{b}}, arXiv e-prints, arXiv:2503.14745

\bibitem[{{Desjacques} {et~al.}(2018){Desjacques}, {Jeong}, \& {Schmidt}}]{Desjacques_2018}
{Desjacques}, V., {Jeong}, D., \& {Schmidt}, F. 2018, \physrep, 733, 1

\bibitem[{Doeser {et~al.}(2024)Doeser, Jamieson, Stopyra, Lavaux, Leclercq, \& Jasche}]{doeserBayesianInferenceInitial2024}
Doeser, L., Jamieson, D., Stopyra, S., {et~al.} 2024, 535, 1258

\bibitem[{Duane {et~al.}(1987)Duane, Kennedy, Pendleton, \& Roweth}]{Duane_1987}
Duane, S., Kennedy, A., Pendleton, B.~J., \& Roweth, D. 1987, Phys. Letters B, 195, 216

\bibitem[{{Eisenstein} {et~al.}(2007){Eisenstein}, {Seo}, {Sirko}, \& {Spergel}}]{Eisenstein_2007}
{Eisenstein}, D.~J., {Seo}, H.-J., {Sirko}, E., \& {Spergel}, D.~N. 2007, \apj, 664, 675

\bibitem[{{Euclid Collaboration} {et~al.}(2025){Euclid Collaboration}, {Aussel}, {Tereno}, {Schirmer}, {Alguero}, {Altieri}, {Balbinot}, {de Boer}, {Casenove}, {Corcho-Caballero}, {Furusawa}, {Furusawa}, {Hudson}, {Jahnke}, {Libet}, {Macias-Perez}, {Masoumzadeh}, {Mohr}, {Odier}, {Scott}, {Vassallo}, {Verdoes Kleijn}, {Zacchei}, {Aghanim}, {Amara}, {Andreon}, {Auricchio}, {Awan}, {Azzollini}, {Baccigalupi}, {Baldi}, {Balestra}, {Bardelli}, {Basset}, {Battaglia}, {Belikov}, {Bender}, {Biviano}, {Bonchi}, {Bonino}, {Branchini}, {Brescia}, {Brinchmann}, {Camera}, {Ca{\~n}as-Herrera}, {Capobianco}, {Carbone}, {Cardone}, {Carretero}, {Casas}, {Castander}, {Castellano}, {Castignani}, {Cavuoti}, {Chambers}, {Cimatti}, {Colodro-Conde}, {Congedo}, {Conselice}, {Conversi}, {Copin}, {Courbin}, {Courtois}, {Cropper}, {Cuby}, {Da Silva}, {da Silva}, {Degaudenzi}, {de Jong}, {De Lucia}, {Di Giorgio}, {Dinis}, {Dolding}, {Dole}, {Douspis}, {Dubath}, {Duncan}, {Dupac}, {Dusini}, {Ealet}, {Escoffier}, {Fabricius}, {Farina},
  {Farinelli}, {Faustini}, {Ferriol}, {Fotopoulou}, {Fourmanoit}, {Frailis}, {Franceschi}, {Franzetti}, {Galeotta}, {George}, {Gillard}, {Gillis}, {Giocoli}, {G{\'o}mez-Alvarez}, {Gracia-Carpio}, {Granett}, {Grazian}, {Grupp}, {Guzzo}, {Gwyn}, {Haugan}, {Herent}, {Hoar}, {Hoekstra}, {Holliman}, {Holmes}, {Hook}, {Hormuth}, {Hornstrup}, {Hudelot}, {Ili{\'c}}, {Jhabvala}, {Joachimi}, {Keih{\"a}nen}, {Kermiche}, {Kiessling}, {Kubik}, {Kuijken}, {K{\"u}mmel}, {Kunz}, {Kurki-Suonio}, {Lahav}, {Le Boulc'h}, {Le Brun}, {Le Mignant}, {Liebing}, {Ligori}, {Lilje}, {Lindholm}, {Lloro}, {Mainetti}, {Maino}, {Maiorano}, {Mansutti}, {Marcin}, {Marggraf}, {Markovic}, {Martinelli}, {Martinet}, {Marulli}, {Massey}, {Maurogordato}, {McCracken}, {Medinaceli}, {Mei}, {Melchior}, {Mellier}, {Meneghetti}, {Merlin}, {Meylan}, {Mora}, {Moresco}, {Morris}, {Moscardini}, {Mourre}, {Nakajima}, {Neissner}, {Nichol}, {Niemi}, {Nightingale}, {Nutma}, {Padilla}, {Paltani}, {Pasian}, {Peacock}, {Pedersen}, {Percival}, {Pettorino}, {Pires},
  {Polenta}, {Pollack}, {Poncet}, {Popa}, {Pozzetti}, {Racca}, {Raison}, {Rebolo}, {Renzi}, {Rhodes}, {Riccio}, {Rix}, {Romelli}, {Roncarelli}, {Rossetti}, {Rusholme}, {Saglia}, {Sakr}, {S{\'a}nchez}, {Sapone}, {Sartoris}, {Sauvage}, {Schewtschenko}, {Schneider}, {Scodeggio}, {Secroun}, {Sefusatti}, \& {Seidel}}]{EUCLID_Q1}
{Euclid Collaboration}, {Aussel}, H., {Tereno}, I., {et~al.} 2025, arXiv e-prints, arXiv:2503.15302

\bibitem[{{Favole} {et~al.}(2025){Favole}, {Kitaura}, {Hadzhiyska}, {Eisenstein}, {Garrison}, \& {Bose}}]{Favole_2025}
{Favole}, G., {Kitaura}, F.-S., {Hadzhiyska}, B., {et~al.} 2025, arXiv e-prints, arXiv:2512.04362, submitted to ApJ; minor corrections

\bibitem[{{Feng} {et~al.}(2016){Feng}, {Chu}, {Seljak}, \& {McDonald}}]{Feng_2016}
{Feng}, Y., {Chu}, M.-Y., {Seljak}, U., \& {McDonald}, P. 2016, MNRAS, 463, 2273

\bibitem[{{Filho} {et~al.}(2015){Filho}, {S{\'a}nchez Almeida}, {Mu{\~n}oz-Tu{\~n}{\'o}n}, {Nuza}, {Kitaura}, \& {He{\ss}}}]{Filho_2015}
{Filho}, M.~E., {S{\'a}nchez Almeida}, J., {Mu{\~n}oz-Tu{\~n}{\'o}n}, C., {et~al.} 2015, \apj, 802, 82

\bibitem[{{Forero S{\'a}nchez} {et~al.}(2024){Forero S{\'a}nchez}, {Kitaura}, {Sinigaglia}, {Coloma-Nadal}, \& {Kneib}}]{Forero_2024}
{Forero S{\'a}nchez}, D., {Kitaura}, F.~S., {Sinigaglia}, F., {Coloma-Nadal}, J.~M., \& {Kneib}, J.~P. 2024, \jcap, 2024, 001

\bibitem[{Forero–Romero {et~al.}(2009)Forero–Romero, Hoffman, Gottlöber, Klypin, \& Yepes}]{Forero_2009}
Forero–Romero, J.~E., Hoffman, Y., Gottlöber, S., Klypin, A., \& Yepes, G. 2009, MNRAS, 396, 1815

\bibitem[{{Fry} \& {Gaztanaga}(1993)}]{Fry_1993}
{Fry}, J.~N. \& {Gaztanaga}, E. 1993, \apj, 413, 447

\bibitem[{{Garc{\'\i}a-Farieta} {et~al.}(2024){Garc{\'\i}a-Farieta}, {Balaguera-Antol{\'\i}nez}, \& {Kitaura}}]{Garcia-Farieta_2024}
{Garc{\'\i}a-Farieta}, J.~E., {Balaguera-Antol{\'\i}nez}, A., \& {Kitaura}, F.-S. 2024, \aap, 690, A27

\bibitem[{{Garrison} {et~al.}(2018){Garrison}, {Eisenstein}, {Ferrer}, {Tinker}, {Pinto}, \& {Weinberg}}]{Garrison_2018}
{Garrison}, L.~H., {Eisenstein}, D.~J., {Ferrer}, D., {et~al.} 2018, \apjs, 236, 43

\bibitem[{{Hahn} \& {Villaescusa-Navarro}(2021)}]{Hahn_2021}
{Hahn}, C. \& {Villaescusa-Navarro}, F. 2021, JCAP, 2021, 029

\bibitem[{{Hahn} {et~al.}(2007){Hahn}, {Porciani}, {Carollo}, \& {Dekel}}]{Hahn_2007}
{Hahn}, O., {Porciani}, C., {Carollo}, C.~M., \& {Dekel}, A. 2007, MNRAS, 375, 489

\bibitem[{{Hamilton}(1998)}]{Hamilton_1998}
{Hamilton}, A.~J.~S. 1998, in Astrophysics and Space Science Library, Vol. 231, The Evolving Universe, ed. D.~{Hamilton}, 185

\bibitem[{{Heavens} \& {Peacock}(1988)}]{Heavens_1988}
{Heavens}, A. \& {Peacock}, J. 1988, MNRAS, 232, 339

\bibitem[{{He{\ss}} {et~al.}(2013){He{\ss}}, {Kitaura}, \& {Gottl{\"o}ber}}]{Hess_2013}
{He{\ss}}, S., {Kitaura}, F.-S., \& {Gottl{\"o}ber}, S. 2013, MNRAS, 435, 2065

\bibitem[{{Hoffman} \& {Gelman}(2011)}]{Hoffman_2011}
{Hoffman}, M.~D. \& {Gelman}, A. 2011, arXiv:1111.4246

\bibitem[{{Horowitz} {et~al.}(2022){Horowitz}, {Dornfest}, {Luki{\'c}}, \& {Harrington}}]{Horowitz_2022}
{Horowitz}, B., {Dornfest}, M., {Luki{\'c}}, Z., \& {Harrington}, P. 2022, \apj, 941, 42

\bibitem[{{Horowitz} {et~al.}(2019{\natexlab{a}}){Horowitz}, {Lee}, {White}, {Krolewski}, \& {Ata}}]{Horowitz_2019b}
{Horowitz}, B., {Lee}, K.-G., {White}, M., {Krolewski}, A., \& {Ata}, M. 2019{\natexlab{a}}, \apj, 887, 61

\bibitem[{Horowitz \& Lukic(2025)}]{horowitzDifferentiableCosmologicalHydrodynamics2025}
Horowitz, B. \& Lukic, Z. 2025 [\eprint{2502.02294}]

\bibitem[{{Horowitz} {et~al.}(2019{\natexlab{b}}){Horowitz}, {Seljak}, \& {Aslanyan}}]{Horowitz_2019}
{Horowitz}, B., {Seljak}, U., \& {Aslanyan}, G. 2019{\natexlab{b}}, JCAP, 2019, 035

\bibitem[{{Horowitz} {et~al.}(2021){Horowitz}, {Zhang}, {Lee}, \& {Kooistra}}]{Horowitz_2021}
{Horowitz}, B., {Zhang}, B., {Lee}, K.-G., \& {Kooistra}, R. 2021, \apj, 906, 110

\bibitem[{{Hui} \& {Gnedin}(1997)}]{Hui_1997}
{Hui}, L. \& {Gnedin}, N.~Y. 1997, MNRAS, 292, 27

\bibitem[{{Ishiyama} {et~al.}(2021){Ishiyama}, {Prada}, {Klypin}, {Sinha}, {Metcalf}, {Jullo}, {Altieri}, {Cora}, {Croton}, {de la Torre}, {Mill{\'a}n-Calero}, {Oogi}, {Ruedas}, \& {Vega-Mart{\'\i}nez}}]{Ishiyama_2021}
{Ishiyama}, T., {Prada}, F., {Klypin}, A.~A., {et~al.} 2021, MNRAS, 506, 4210

\bibitem[{{Ivanov} {et~al.}(2025){Ivanov}, {Obuljen}, {Cuesta-Lazaro}, \& {Toomey}}]{Ivanov_2025}
{Ivanov}, M.~M., {Obuljen}, A., {Cuesta-Lazaro}, C., \& {Toomey}, M.~W. 2025, \prd, 111, 063548

\bibitem[{{Jasche} \& {Kitaura}(2010)}]{Jasche_2010}
{Jasche}, J. \& {Kitaura}, F.-S. 2010, MNRAS, 407, 29

\bibitem[{{Jasche} {et~al.}(2010){Jasche}, {Kitaura}, {Li}, \& {En{\ss}lin}}]{Jasche_2010b}
{Jasche}, J., {Kitaura}, F.~S., {Li}, C., \& {En{\ss}lin}, T.~A. 2010, \mnras, 409, 355

\bibitem[{{Jasche} \& {Lavaux}(2019)}]{Jasche_2019}
{Jasche}, J. \& {Lavaux}, G. 2019, \aap, 625, A64

\bibitem[{Jasche {et~al.}(2015)Jasche, Leclercq, \& Wandelt}]{jaschePresentCosmicStructure2015}
Jasche, J., Leclercq, F., \& Wandelt, B. 2015, 2015, 036

\bibitem[{{Jasche} \& {Wandelt}(2013)}]{Jasche_2013}
{Jasche}, J. \& {Wandelt}, B.~D. 2013, MNRAS, 432, 894

\bibitem[{{Kaiser}(1984)}]{Kaiser_1984}
{Kaiser}, N. 1984, \apjl, 284, L9

\bibitem[{{Kaiser}(1987)}]{Kaiser_1987}
{Kaiser}, N. 1987, MNRAS, 227, 1

\bibitem[{{Kitaura}(2013)}]{Kitaura_2013a}
{Kitaura}, F.-S. 2013, MNRAS: Letters, 429, L84

\bibitem[{{Kitaura} {et~al.}(2012{\natexlab{a}}){Kitaura}, {Angulo}, {Hoffman}, \& {Gottl{\"o}ber}}]{Kitaura_2012c}
{Kitaura}, F.-S., {Angulo}, R.~E., {Hoffman}, Y., \& {Gottl{\"o}ber}, S. 2012{\natexlab{a}}, MNRAS, 425, 2422

\bibitem[{{Kitaura} {et~al.}(2016{\natexlab{a}}){Kitaura}, {Ata}, {Angulo}, {Chuang}, {Rodr{\'\i}guez-Torres}, {Monteagudo}, {Prada}, \& {Yepes}}]{Kitaura_2016a}
{Kitaura}, F.-S., {Ata}, M., {Angulo}, R.~E., {et~al.} 2016{\natexlab{a}}, MNRAS: Letters, 457, L113

\bibitem[{{Kitaura} {et~al.}(2021){Kitaura}, {Ata}, {Rodr{\'\i}guez-Torres}, {Hern{\'a}ndez-S{\'a}nchez}, {Balaguera-Antol{\'\i}nez}, \& {Yepes}}]{Kitaura_2021}
{Kitaura}, F.-S., {Ata}, M., {Rodr{\'\i}guez-Torres}, S.~A., {et~al.} 2021, MNRAS, 502, 3456

\bibitem[{{Kitaura} {et~al.}(2022){Kitaura}, {Balaguera-Antol{\'\i}nez}, {Sinigaglia}, \& {Pellejero-Ib{\'a}{\~n}ez}}]{Kitaura_2022}
{Kitaura}, F.-S., {Balaguera-Antol{\'\i}nez}, A., {Sinigaglia}, F., \& {Pellejero-Ib{\'a}{\~n}ez}, M. 2022, MNRAS, 512, 2245

\bibitem[{{Kitaura} \& {En{\ss}lin}(2008)}]{Kitaura_2008}
{Kitaura}, F.-S. \& {En{\ss}lin}, T.~A. 2008, MNRAS, 389, 497

\bibitem[{{Kitaura} {et~al.}(2012{\natexlab{b}}){Kitaura}, {Erdo{\v g}du}, {Nuza}, {Khalatyan}, {Angulo}, {Hoffman}, \& {Gottl{\"o}ber}}]{Kitaura_2012a}
{Kitaura}, F.-S., {Erdo{\v g}du}, P., {Nuza}, S.~E., {et~al.} 2012{\natexlab{b}}, MNRAS: Letters, 427, L35

\bibitem[{{Kitaura} {et~al.}(2012{\natexlab{c}}){Kitaura}, {Gallerani}, \& {Ferrara}}]{Kitaura_2012}
{Kitaura}, F.-S., {Gallerani}, S., \& {Ferrara}, A. 2012{\natexlab{c}}, MNRAS, 420, 61

\bibitem[{{Kitaura} {et~al.}(2015){Kitaura}, {Gil-Mar{\'\i}n}, {Sc{\'o}ccola}, {Chuang}, {M{\"u}ller}, {Yepes}, \& {Prada}}]{Kitaura_2015}
{Kitaura}, F.-S., {Gil-Mar{\'\i}n}, H., {Sc{\'o}ccola}, C.~G., {et~al.} 2015, MNRAS, 450, 1836

\bibitem[{Kitaura \& He{\ss}(2013)}]{Kitaura_2013}
Kitaura, F.-S. \& He{\ss}, S. 2013, MNRAS: Letters, 435, L78

\bibitem[{{Kitaura} {et~al.}(2010){Kitaura}, {Jasche}, \& {Metcalf}}]{Kitaura_2010}
{Kitaura}, F.-S., {Jasche}, J., \& {Metcalf}, R.~B. 2010, MNRAS, 403, 589

\bibitem[{{Kitaura} {et~al.}(2016{\natexlab{b}}){Kitaura}, {Rodr{\'\i}guez-Torres}, {Chuang}, {Zhao}, {Prada}, {Gil-Mar{\'\i}n}, {Guo}, {Yepes}, {Klypin}, {Sc{\'o}ccola}, {Tinker}, {McBride}, {Reid}, {S{\'a}nchez}, {Salazar-Albornoz}, {Grieb}, {Vargas-Magana}, {Cuesta}, {Neyrinck}, {Beutler}, {Comparat}, {Percival}, \& {Ross}}]{Kitaura_2016}
{Kitaura}, F.-S., {Rodr{\'\i}guez-Torres}, S., {Chuang}, C.-H., {et~al.} 2016{\natexlab{b}}, MNRAS, 456, 4156

\bibitem[{{Kitaura} {et~al.}(2024){Kitaura}, {Sinigaglia}, {Balaguera-Antol{\'\i}nez}, \& {Favole}}]{Kitaura_2024}
{Kitaura}, F.-S., {Sinigaglia}, F., {Balaguera-Antol{\'\i}nez}, A., \& {Favole}, G. 2024, \aap, 683, A215

\bibitem[{Kitaura {et~al.}(2014)Kitaura, Yepes, \& Prada}]{Kitaura_2014}
Kitaura, F.-S., Yepes, G., \& Prada, F. 2014, MNRAS: Letters, 439, L21

\bibitem[{{Klypin} \& {Prada}(2018)}]{Klypin_2018}
{Klypin}, A. \& {Prada}, F. 2018, MNRAS, 478, 4602

\bibitem[{{Kodi Ramanah} {et~al.}(2020){Kodi Ramanah}, {Charnock}, {Villaescusa-Navarro}, \& {Wandelt}}]{Kodi_2020}
{Kodi Ramanah}, D., {Charnock}, T., {Villaescusa-Navarro}, F., \& {Wandelt}, B.~D. 2020, \mnras, 495, 4227

\bibitem[{Lavaux \& Jasche(2016)}]{lavauxUnmaskingMaskedUniverse2016}
Lavaux, G. \& Jasche, J. 2016, 455, 3169

\bibitem[{Lavaux {et~al.}(2019)Lavaux, Jasche, \& Leclercq}]{lavauxSystematicfreeInferenceCosmic2019}
Lavaux, G., Jasche, J., \& Leclercq, F. 2019, Systematic-Free Inference of the Cosmic Matter Density Field from {{SDSS3-BOSS}} Data

\bibitem[{{Levi} {et~al.}(2013){Levi}, {Bebek}, {Beers}, {Blum}, {Cahn}, {Eisenstein}, {Flaugher}, {Honscheid}, {Kron}, {Lahav}, {McDonald}, {Roe}, {Schlegel}, \& {representing the DESI collaboration}}]{DESI}
{Levi}, M., {Bebek}, C., {Beers}, T., {et~al.} 2013 [\eprint[arXiv]{1308.0847}]

\bibitem[{{Li} {et~al.}(2024){Li}, {Modi}, {Jamieson}, {Zhang}, {Lu}, {Feng}, {Lanusse}, \& {Greengard}}]{Li_2024}
{Li}, Y., {Modi}, C., {Jamieson}, D., {et~al.} 2024, \apjs, 270, 36

\bibitem[{{McDonald} {et~al.}(2000){McDonald}, {Miralda-Escud{\'e}}, {Rauch}, {Sargent}, {Barlow}, {Cen}, \& {Ostriker}}]{McDonald_2000}
{McDonald}, P., {Miralda-Escud{\'e}}, J., {Rauch}, M., {et~al.} 2000, \apj, 543, 1

\bibitem[{McDonald \& Roy(2009)}]{McDonald_2009}
McDonald, P. \& Roy, A. 2009, JCAP, 2009, 020

\bibitem[{Modi {et~al.}(2021)Modi, Lanusse, \& Seljak}]{Modi_2021}
Modi, C., Lanusse, F., \& Seljak, U. 2021, Astronomy and Computing, 37, 100505

\bibitem[{{Monaco} \& {Efstathiou}(1999)}]{Monaco_1999}
{Monaco}, P. \& {Efstathiou}, G. 1999, MNRAS, 308, 763

\bibitem[{{Neyrinck}(2016)}]{Neyrinck_2016}
{Neyrinck}, M.~C. 2016, MNRAS, 455, L11

\bibitem[{{Neyrinck} {et~al.}(2014){Neyrinck}, {Arag{\'o}n-Calvo}, {Jeong}, \& {Wang}}]{Neyrinck_2014}
{Neyrinck}, M.~C., {Arag{\'o}n-Calvo}, M.~A., {Jeong}, D., \& {Wang}, X. 2014, MNRAS, 441, 646

\bibitem[{{Nusser} \& {Dekel}(1992)}]{Nusser_1992}
{Nusser}, A. \& {Dekel}, A. 1992, \apj, 391, 443

\bibitem[{{Nuza} {et~al.}(2014){Nuza}, {Kitaura}, {He{\ss}}, {Libeskind}, \& {M{\"u}ller}}]{Nuza_2014}
{Nuza}, S.~E., {Kitaura}, F.-S., {He{\ss}}, S., {Libeskind}, N.~I., \& {M{\"u}ller}, V. 2014, MNRAS, 445, 988

\bibitem[{Olex {et~al.}(2025)Olex, Hellwing, \& Knebe}]{olexUniversalPhysicallyMotivated2025}
Olex, E., Hellwing, W.~A., \& Knebe, A. 2025, 696, A142

\bibitem[{{Paillas} {et~al.}(2025){Paillas}, {Ding}, {Chen}, {Seo}, {Padmanabhan}, {de Mattia}, {Ross}, {Nadathur}, {Howlett}, {Aguilar}, {Ahlen}, {Alves}, {Andrade}, {Brooks}, {Buckley-Geer}, {Burtin}, {Chen}, {Claybaugh}, {Cole}, {Dawson}, {de la Macorra}, {Dey}, {Doel}, {Fanning}, {Ferraro}, {Forero-Romero}, {Garcia-Quintero}, {Gazta{\~n}aga}, {Gil-Mar{\'\i}n}, {Gontcho}, {Gutierrez}, {Hahn}, {Hanif}, {Honscheid}, {Ishak}, {Kehoe}, {Kremin}, {Landriau}, {Le Guillou}, {Levi}, {Manera}, {Martini}, {Medina-Varela}, {Meisner}, {Mena-Fern{\'a}ndez}, {Miquel}, {Moustakas}, {Mueller}, {Mu{\~n}oz-Guti{\'e}rrez}, {Myers}, {Newman}, {Nie}, {Niz}, {Palanque-Delabrouille}, {Percival}, {Poppett}, {Prada}, {P{\'e}rez-Fern{\'a}ndez}, {Rashkovetskyi}, {Rezaie}, {Rosado-Marin}, {Rossi}, {Ruggeri}, {Sanchez}, {Saulder}, {Schlafly}, {Schlegel}, {Schubnell}, {Sprayberry}, {Tarl{\'e}}, {Valcin}, {Vargas-Maga{\~n}a}, {Yu}, {Yuan}, {Zhou}, \& {Zou}}]{Paillas_2025}
{Paillas}, E., {Ding}, Z., {Chen}, X., {et~al.} 2025, JCAP, 2025, 142

\bibitem[{{Peebles}(1980)}]{Peebles_1980}
{Peebles}, P.~J.~E. 1980, {The large-scale structure of the universe} (Princeton University Press)

\bibitem[{Pellejero-Ibañez {et~al.}(2020)Pellejero-Ibañez, Balaguera-Antolínez, Kitaura, Angulo, Yepes, Chuang, Reyes-Peraza, Autefage, Vakili, \& Zhao}]{Pellejero_2020}
Pellejero-Ibañez, M., Balaguera-Antolínez, A., Kitaura, F.-S., {et~al.} 2020, MNRAS, 493, 586

\bibitem[{Phan {et~al.}(2019)Phan, Pradhan, \& Jankowiak}]{phanComposableEffectsFlexible2019}
Phan, D., Pradhan, N., \& Jankowiak, M. 2019 [\eprint{1912.11554}]

\bibitem[{Porqueres {et~al.}(2019)Porqueres, Jasche, Lavaux, \& Enßlin}]{porqueresInferringHighredshiftLargescale2019}
Porqueres, N., Jasche, J., Lavaux, G., \& Enßlin, T. 2019, 630, A151

\bibitem[{{Rodr{\'\i}guez-Torres} {et~al.}(2016){Rodr{\'\i}guez-Torres}, {Chuang}, {Prada}, {Guo}, {Klypin}, {Behroozi}, {Hahn}, {Comparat}, {Yepes}, {Montero-Dorta}, {Brownstein}, {Maraston}, {McBride}, {Tinker}, {Gottl{\"o}ber}, {Favole}, {Shu}, {Kitaura}, {Bolton}, {Scoccimarro}, {Samushia}, {Schlegel}, {Schneider}, \& {Thomas}}]{Torres_2016}
{Rodr{\'\i}guez-Torres}, S.~A., {Chuang}, C.-H., {Prada}, F., {et~al.} 2016, MNRAS, 460, 1173

\bibitem[{{Saslaw} \& {Hamilton}(1984)}]{Saslaw_1984}
{Saslaw}, W.~C. \& {Hamilton}, A.~J.~S. 1984, \apj, 276, 13

\bibitem[{{Schmittfull} {et~al.}(2019){Schmittfull}, {Simonovi{\'c}}, {Assassi}, \& {Zaldarriaga}}]{Schmittfull_2019}
{Schmittfull}, M., {Simonovi{\'c}}, M., {Assassi}, V., \& {Zaldarriaga}, M. 2019, \prd, 100, 043514

\bibitem[{{Seljak}(2012)}]{Seljak_2012}
{Seljak}, U. 2012, JCAP, 2012, 004

\bibitem[{Sheth(1995)}]{Sheth_1995}
Sheth, R.~K. 1995, MNRAS, 274, 213

\bibitem[{{Sheth} \& {Lemson}(1999)}]{Sheth_1999}
{Sheth}, R.~K. \& {Lemson}, G. 1999, MNRAS, 304, 767

\bibitem[{{Shirasaki} {et~al.}(2021){Shirasaki}, {Sugiyama}, {Takahashi}, \& {Kitaura}}]{Shirasaki_2021}
{Shirasaki}, M., {Sugiyama}, N.~S., {Takahashi}, R., \& {Kitaura}, F.-S. 2021, \prd, 103, 023506

\bibitem[{{Sinigaglia} {et~al.}(2022){Sinigaglia}, {Kitaura}, {Balaguera-Antol{\'\i}nez}, {Shimizu}, {Nagamine}, {S{\'a}nchez-Benavente}, \& {Ata}}]{Sinigaglia_2022}
{Sinigaglia}, F., {Kitaura}, F.-S., {Balaguera-Antol{\'\i}nez}, A., {et~al.} 2022, \apj, 927, 230

\bibitem[{Sinigaglia {et~al.}(2021)Sinigaglia, Kitaura, Balaguera-Antolínez, Nagamine, Ata, Shimizu, \& Sánchez-Benavente}]{Sinigaglia_2021}
Sinigaglia, F., Kitaura, F.-S., Balaguera-Antolínez, A., {et~al.} 2021, ApJ, 921, 66

\bibitem[{{Sinigaglia} {et~al.}(2024{\natexlab{a}}){Sinigaglia}, {Kitaura}, {Nagamine}, \& {Oku}}]{Sinigaglia_2024b}
{Sinigaglia}, F., {Kitaura}, F.-S., {Nagamine}, K., \& {Oku}, Y. 2024{\natexlab{a}}, ApJ: Letters, 971, L22

\bibitem[{{Sinigaglia} {et~al.}(2024{\natexlab{b}}){Sinigaglia}, {Kitaura}, {Nagamine}, {Oku}, \& {Balaguera-Antol{\'\i}nez}}]{Sinigaglia_2024}
{Sinigaglia}, F., {Kitaura}, F.-S., {Nagamine}, K., {Oku}, Y., \& {Balaguera-Antol{\'\i}nez}, A. 2024{\natexlab{b}}, \aap, 682, A21

\bibitem[{{Somerville} {et~al.}(2001){Somerville}, {Lemson}, {Sigad}, {Dekel}, {Kauffmann}, \& {White}}]{Somerville_2001}
{Somerville}, R.~S., {Lemson}, G., {Sigad}, Y., {et~al.} 2001, MNRAS, 320, 289

\bibitem[{Stadler {et~al.}(2023)Stadler, Schmidt, \& Reinecke}]{stadlerCosmologyInferenceField2023}
Stadler, J., Schmidt, F., \& Reinecke, M. 2023, 2023, 069

\bibitem[{{Takada} {et~al.}(2014){Takada}, {Ellis}, {Chiba}, {Greene}, {Aihara}, {Arimoto}, {Bundy}, {Cohen}, {Dor{\'e}}, {Graves}, {Gunn}, {Heckman}, {Hirata}, {Ho}, {Kneib}, {Le F{\`e}vre}, {Lin}, {More}, {Murayama}, {Nagao}, {Ouchi}, {Seiffert}, {Silverman}, {Sodr{\'e}}, {Spergel}, {Strauss}, {Sugai}, {Suto}, {Takami}, \& {Wyse}}]{Takada_2014}
{Takada}, M., {Ellis}, R.~S., {Chiba}, M., {et~al.} 2014, \pasj, 66, R1

\bibitem[{{Tassev} {et~al.}(2013){Tassev}, {Zaldarriaga}, \& {Eisenstein}}]{Tassev_2013}
{Tassev}, S., {Zaldarriaga}, M., \& {Eisenstein}, D.~J. 2013, JCAP, 6, 036

\bibitem[{{Tegmark} {et~al.}(2004){Tegmark}, {Blanton}, {Strauss}, {Hoyle}, {Schlegel}, {Scoccimarro}, {Vogeley}, {Weinberg}, {Zehavi}, {Berlind}, {Budavari}, {Connolly}, {Eisenstein}, {Finkbeiner}, {Frieman}, {Gunn}, {Hamilton}, {Hui}, {Jain}, {Johnston}, {Kent}, {Lin}, {Nakajima}, {Nichol}, {Ostriker}, {Pope}, {Scranton}, {Seljak}, {Sheth}, {Stebbins}, {Szalay}, {Szapudi}, {Verde}, {Xu}, {Annis}, {Bahcall}, {Brinkmann}, {Burles}, {Castander}, {Csabai}, {Loveday}, {Doi}, {Fukugita}, {Gott}, {Hennessy}, {Hogg}, {Ivezi{\'c}}, {Knapp}, {Lamb}, {Lee}, {Lupton}, {McKay}, {Kunszt}, {Munn}, {O'Connell}, {Peoples}, {Pier}, {Richmond}, {Rockosi}, {Schneider}, {Stoughton}, {Tucker}, {Vanden Berk}, {Yanny}, {York}, \& {SDSS Collaboration}}]{Tegmark_2004}
{Tegmark}, M., {Blanton}, M.~R., {Strauss}, M.~A., {et~al.} 2004, \apj, 606, 702

\bibitem[{{Vos-Gin{\'e}s} {et~al.}(2024){Vos-Gin{\'e}s}, {Avila}, {Gonzalez-Perez}, \& {Yepes}}]{2024MNRAS.530.3458V}
{Vos-Gin{\'e}s}, B., {Avila}, S., {Gonzalez-Perez}, V., \& {Yepes}, G. 2024, \mnras, 530, 3458

\bibitem[{{Wang} {et~al.}(2014){Wang}, {Mo}, {Yang}, {Jing}, \& {Lin}}]{Wang_2014}
{Wang}, H., {Mo}, H.~J., {Yang}, X., {Jing}, Y.~P., \& {Lin}, W.~P. 2014, \apj, 794, 94

\bibitem[{{Wang} {et~al.}(2012){Wang}, {Mo}, {Yang}, \& {van den Bosch}}]{Wang_2012}
{Wang}, H., {Mo}, H.~J., {Yang}, X., \& {van den Bosch}, F.~C. 2012, MNRAS, 420, 1809

\bibitem[{{Wang} {et~al.}(2013){Wang}, {Mo}, {Yang}, \& {Van den Bosch}}]{Wang_2013}
{Wang}, H., {Mo}, H.~J., {Yang}, X., \& {Van den Bosch}, F.~C. 2013, \apj, 772, 63

\bibitem[{{Wang} {et~al.}(2022){Wang}, {Zhai}, {Alavi}, {Massara}, {Pisani}, {Benson}, {Hirata}, {Samushia}, {Weinberg}, {Colbert}, {Dor{\'e}}, {Eifler}, {Heinrich}, {Ho}, {Krause}, {Padmanabhan}, {Spergel}, \& {Teplitz}}]{Wang_2022}
{Wang}, Y., {Zhai}, Z., {Alavi}, A., {et~al.} 2022, \apj, 928, 1

\bibitem[{{Werner} \& {Porciani}(2020)}]{Werner_2020}
{Werner}, K.~F. \& {Porciani}, C. 2020, MNRAS, 492, 1614

\bibitem[{{Yahil} {et~al.}(1991){Yahil}, {Strauss}, {Davis}, \& {Huchra}}]{Yahil_1991}
{Yahil}, A., {Strauss}, M.~A., {Davis}, M., \& {Huchra}, J.~P. 1991, \apj, 372, 380

\bibitem[{{Zaroubi} {et~al.}(1995){Zaroubi}, {Hoffman}, {Fisher}, \& {Lahav}}]{Zaroubi_1995}
{Zaroubi}, S., {Hoffman}, Y., {Fisher}, K.~B., \& {Lahav}, O. 1995, \apj, 449, 446

\bibitem[{{Zel'dovich}(1970)}]{Zeldovich_1970}
{Zel'dovich}, Y.~B. 1970, \aap, 5, 84

\bibitem[{{Zhao} {et~al.}(2024){Zhao}, {Huang}, {He}, {Montero-Camacho}, {Liu}, {Renard}, {Tang}, {Verdier}, {Xu}, {Yang}, {Yu}, {Zhang}, {Zhao}, {Zhou}, {He}, {Kneib}, {Li}, {Li}, {Wang}, {Xianyu}, {Zhang}, {Gsponer}, {Li}, {Rocher}, {Zou}, {Tan}, {Huang}, {Wang}, {Li}, {Rombach}, {Dong}, {Forero-Sanchez}, {Shan}, {Wang}, {Li}, {Zhai}, {Wang}, {Zhao}, {Shi}, {Mao}, {Huang}, {Guo}, \& {Cai}}]{MUST_2024}
{Zhao}, C., {Huang}, S., {He}, M., {et~al.} 2024, arXiv:2411.07970

\end{thebibliography}

\begin{appendix}

\section{White-noise parametrization}
\label{app:white}

In the forward model we have adopted a white-noise parametrization, \(\vect{\nu}\), of i.i.d. standard normal variates, which we transform into the linear density field through the linear operator
\begin{equation}
    \mathcal{M}_\delta(\vect{\nu}) \equiv 
    \vect{\delta}_0 =
    (\Delta x)^{-3}\,
    \mathsf{F}^{-1}\sqrt{\mathsf{P}}\;\mathsf{F}\,\vect{\nu},
\end{equation}
where \(\mathsf{F}\) is the discrete Fourier-transform (DFT) linear operator on an \(N^{3}\) mesh of side length \(L\),  given by 
\begin{align}
    \mathsf{F}_{nm} &= (\Delta x)^3 \exp \left(-i \vect{k}_n \cdot \vect{x}_m\right), \\
    \mathsf{F}^{-1}_{nm} &= L^{-3} \exp \left(i \vect{k}_m \cdot \vect{x}_n\right) \, .
\end{align}
Here \(\vect{x}_n\) represents the Cartesian direction vector in commoving space at a voxel indexed by \(n\). The same holds for the wave-vector \(\vect{k}_n\) in the Fourier-transformed space. We have defined the diagonal linear matter power spectrum matrix 
 \(\mathsf{P}_{nm}\equiv\delta^{\mathrm K}_{nm}\,P_{\mathrm{lin}}(|\vect k_{n}|)\). 
The density contrast, \(\vect{\delta}_0\), is then Gaussian-distributed with zero mean and covariance:
\begin{equation}
    \mathsf{C} = (\Delta x)^{-3} \mathsf{F}^{-1} \mathsf{P} \mathsf{F} \ .
\end{equation}
In principle, one can choose to parametrize the field either in terms of \(\vect{\delta}_0\) directly or through the underlying white noise \(\vect{\nu}\).
Sampling \(\vect{\delta}_0\) directly removes the linear transformation from the forward model but the fast Fourier transforms must still be performed at every Markov chain Monte Carlo step to evaluate the negative log prior and to draw new field realisations.
We have tested both parametrizations and did not observe significant differences in computational cost or sampling behaviour. Nonetheless, we favour the white-noise parametrization \(\vect{\nu}\) for its simpler implementation. 

Another option is to parametrize the field directly in Fourier space. Defining
\begin{equation}
    \hat{\vect{u}} = \sqrt{\frac{2 N^3}{L^6}} \, \mathsf{F} \vect{\nu},
\end{equation}
the components of \(\hat{\vect{u}}\) are i.i.d standard normal variates for both real and imaginary parts. The density contrast is then given by
\begin{equation}
    \vect{\delta}_0 = \sqrt{\frac{L^3}{2}} \, \mathsf{F}^{-1} \sqrt{\mathsf{P}} \, \hat{\vect{u}} \ .
\end{equation}
With this parametrization every parameter corresponds to a definite scale/wavenumber. 
We expected a different chain sampling behaviour, especially when using diagonal mass matrix adaptation during burn-in, but we saw no significant differences with respect to the white noise parametrization. 
We note that this parametrization requires the careful construction of a 
Hermitian-packed Fourier (\(N,N,N/2+1\))-shaped array produced by a 3D real-to-complex fast Fourier transform out of \(N^3\) random real numbers, with explicit handling of the array structure and Nyquist planes.

A third option we considered is to separate each Fourier mode into amplitude and
phase,
\(
    \hat{\vect{\delta}}_{0} = |\vect{\delta}_{0}|\exp(i\vect{\varphi}) \ .
\)
The parameters to be sampled are then split into:
\begin{itemize}
    \item \(N^3/2-4\) phases uniformly distributed in \([0, 2\pi)\)
    \item \(N^3/2-4\) amplitudes following a Rayleigh distribution.
    \item \(8\) Nyquist modes following a Gaussian distribution. 
\end{itemize}This parametrization may be advantageous if we want to fix the amplitudes to the theoretical linear matter power spectrum, effectively reducing the parameter space in half. But again, for our work, we did not come across any obvious advantages of using this approach over the more straightforward white-noise parametrization.

\section{Nonlocal bias and cosmic-web relation}
\label{app:types}

Let us consider a scalar field \(\eta\), which could be, for example, the matter overdensity field \(\eta = \delta\) or the gravitational potential \(\eta = \Phi\). We begin by constructing the Hessian matrix of this field, defined as 
\begin{equation}
\mathcal{H}_{jk} = \frac{\partial^2 \eta}{\partial x_j \partial x_k} = \eta_{,jk}\,,
\end{equation}
which encodes the second derivatives of \(\eta\) with respect to spatial coordinates.

We then compute the eigenvalues of the Hessian, denoted by \(\lambda^{(1)} \geq \lambda^{(2)} \geq \lambda^{(3)}\). These eigenvalues characterize the local curvature of the field in each principal direction. From them, we define the three rotationally invariant scalar quantities, or invariants, of the Hessian:
\begin{itemize}
    \item \(I_1 = \lambda^{(1)} + \lambda^{(2)} + \lambda^{(3)}\), the trace of the Hessian,
    \item \(I_2 = \lambda^{(1)}\lambda^{(2)} + \lambda^{(1)}\lambda^{(3)} + \lambda^{(2)}\lambda^{(3)}\), the sum of principal minors,
    \item \(I_3 = \lambda^{(1)}\lambda^{(2)}\lambda^{(3)}\), the determinant of the Hessian.
\end{itemize}
According to \citet{Kitaura_2022}, the invariants of the Hessian matrix can be directly linked to different regions of the cosmic-web. For simplicity, and without loss of generality, this classification can be expressed for a threshold \(\lambda_{\rm th} = 0\) as follows:
\begin{itemize}
    \item Knots: \quad \(I_3 > 0\), \quad \(I_2 > 0\), \quad \(I_1 > \lambda^{(1)}\)
    \item Filaments:
    \(I_3 < 0\), \quad \(I_2 < 0\), \quad \text{or} 
    
    \quad
         \(I_3 < 0\), \quad \(I_2 > 0\), \quad \(\lambda^{(3)} < I_1 < \lambda^{(3)} - \frac{\lambda^{(2)} \lambda^{(3)}}{\lambda^{(1)}}\)
    \item Sheets:
         \(I_3 > 0\), \quad \(I_2 < 0\), \quad \text{or}
         
        \quad \(I_3 < 0\), \quad \(I_2 > 0\), \quad \(\lambda^{(1)} - \frac{\lambda^{(2)} \lambda^{(3)}}{\lambda^{(1)}} < I_1 < \lambda^{(1)}\)
    \item Voids: \quad \(I_3 < 0\), \quad \(I_2 > 0\), \quad \(I_1 < \lambda^{(1)}\).

\end{itemize}
Moreover, these invariants can be directly related to the nonlocal bias operators commonly used in perturbation theory. Specifically for $\eta=\Phi$, i.e., long-range nonlocal bias:
\begin{itemize}
    \item \(\delta = I_1\), \quad the local density contrast,
    \item \(s^2 = \frac{2}{3}I_1^2 - 2I_2\), \quad the tidal shear squared,
    \item \(s^3 = -I_1 I_2 + 3 I_3 + \frac{2}{9}I_1^3\), \quad the cubic tidal bias term.
\end{itemize}
This formulation thus bridges cosmic-web morphology with perturbative bias theory, showing that both short-range (\(\eta = \delta\)) and long-range (\(\eta = \Phi\)) nonlocal bias terms can be systematically described in terms of the same invariant-based framework. We refer to the cosmic-web classification based on \(\eta = \delta\) as the $\delta$-web, and when based on \(\eta = \Phi\), as the \(\Phi\)-web.

\section{Numerical results of \ttwo and \tthree}
\label{app:results}

\begin{figure*}[p]
\centering
\includegraphics[width=0.8\textwidth]{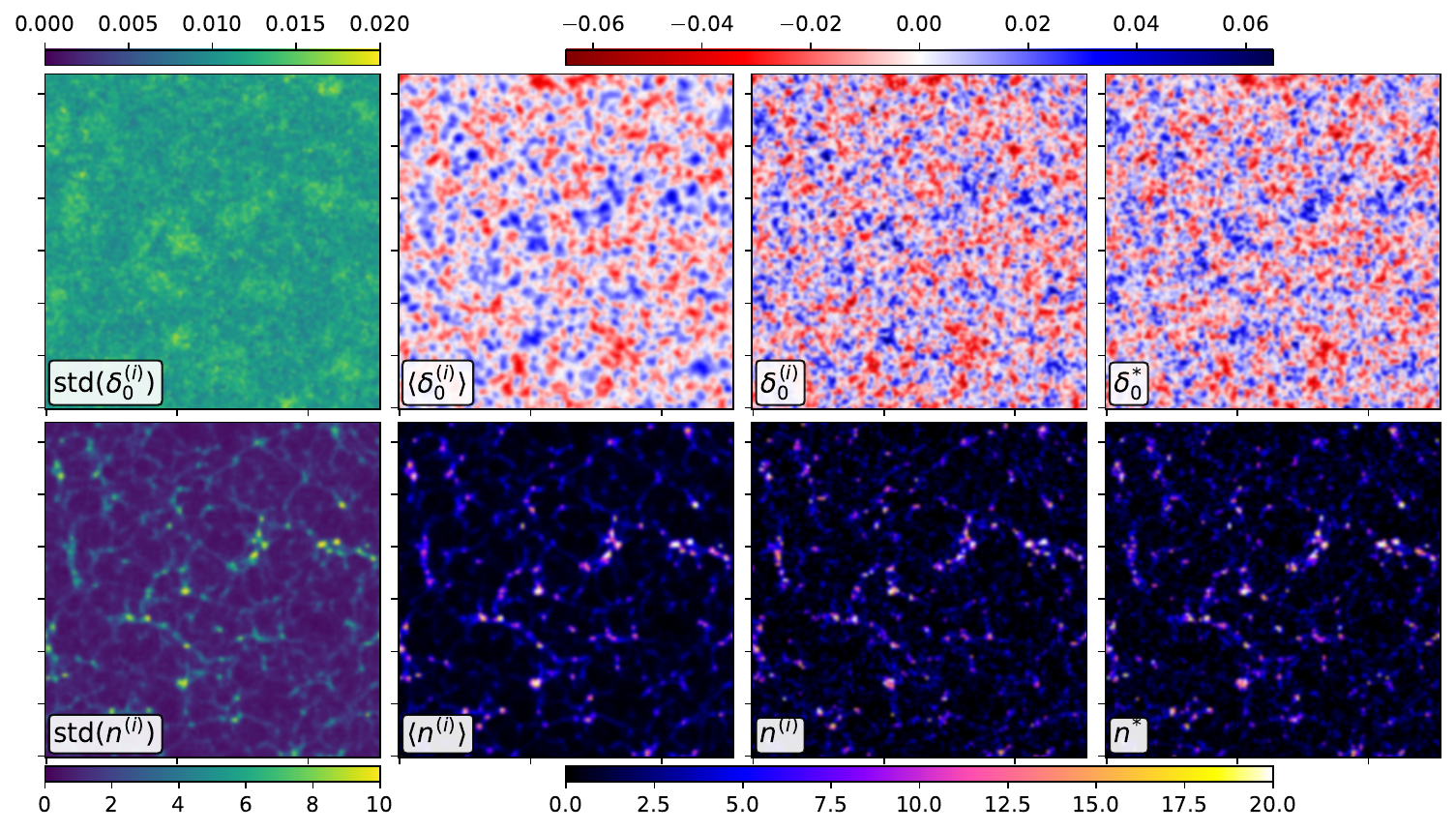}
    \caption{\ttwo: Analogous to Fig.~\ref{fig:maps}.}
          \label{fig:maps5}
\end{figure*}

\begin{figure*}[p]
\centering
\includegraphics[width=1\textwidth]{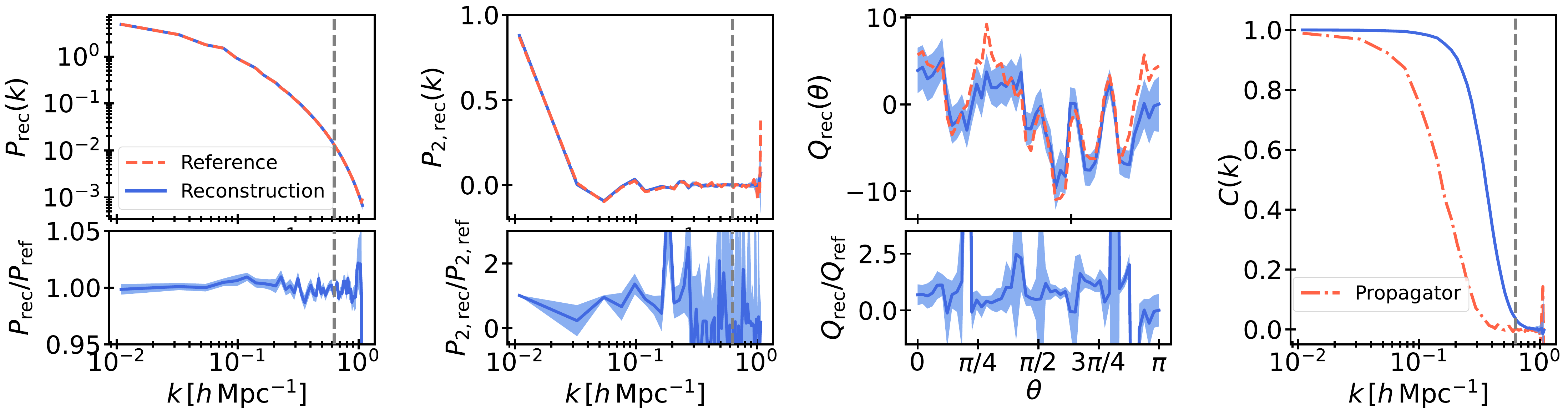}
    \caption{\ttwo:  Initial conditions. Analogous to Fig.~\ref{fig:din_stats}.}
      \label{fig:din_stats5}
\end{figure*}

\begin{figure*}[p]
\centering
\includegraphics[width=1\textwidth]{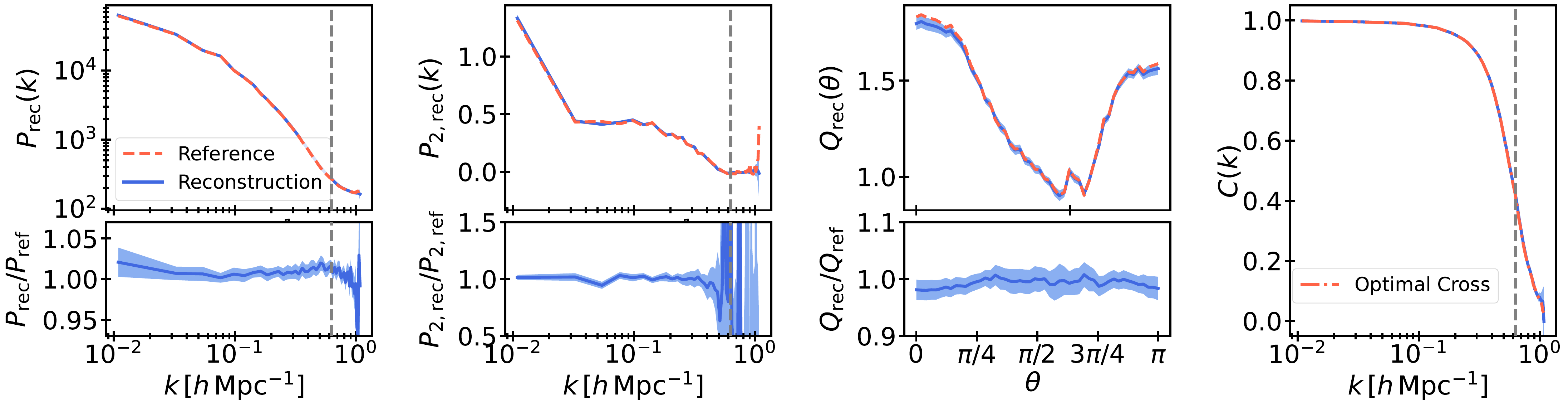}
    \caption{\ttwo:  Final conditions.  Analogous to Fig.~\ref{fig:ntr_stats}.}
          \label{fig:ntr_stats5}
\end{figure*}

Here we present the summary statistics resulting from \ttwo and \tthree. 
The case of \ttwo\ -- when the resolution is doubled to 5~\mpch{} with respect to \tone\ -- yields similarly accurate results across all summary statistics (see 
Figs. \ref{fig:maps5}, \ref{fig:din_stats5}, and  \ref{fig:ntr_stats5}).

\begin{figure*}[p]
\centering
\includegraphics[width=0.8\textwidth]{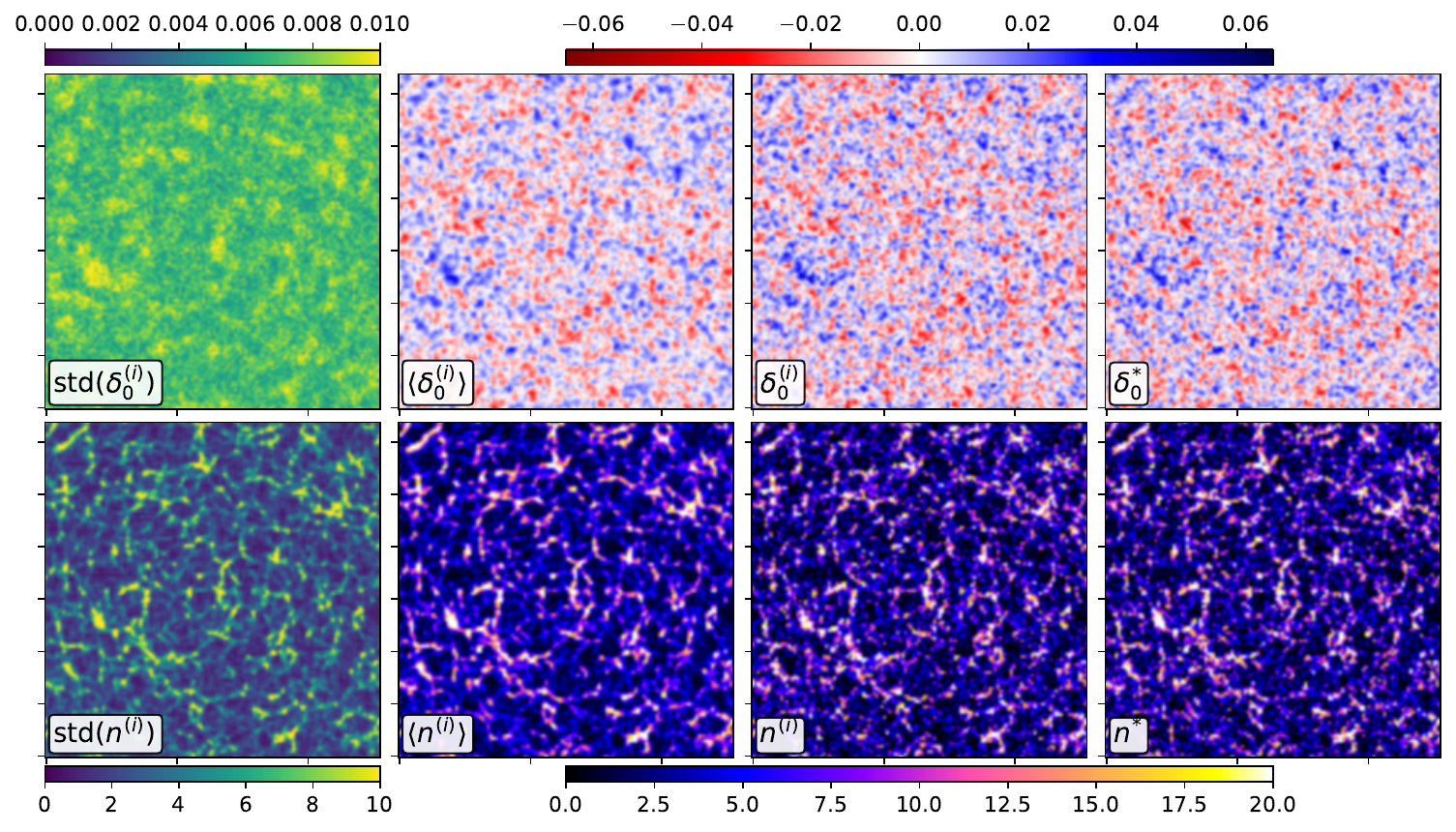}
    \caption{\tthree: Analogous to Fig.~\ref{fig:maps}.}
          \label{fig:maps8}
\end{figure*}

\begin{figure*}[p]
\centering
\includegraphics[width=1\textwidth]{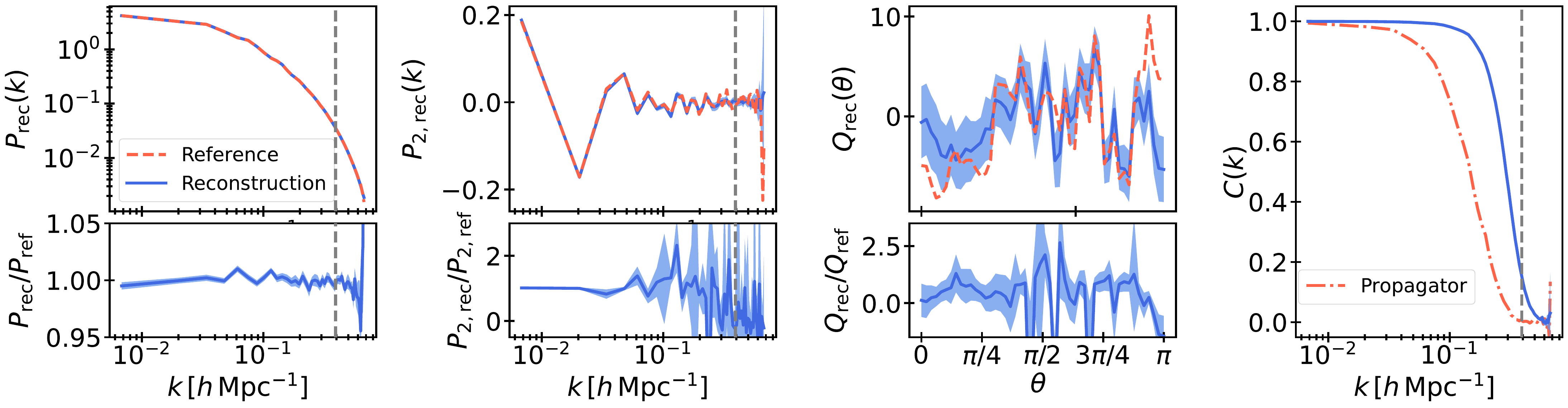}
    \caption{\tthree:  Initial conditions. Analogous to Fig.~\ref{fig:din_stats}.}
\label{fig:din_stats5HIC}
\end{figure*}

\begin{figure*}[p]
\centering
\includegraphics[width=1\textwidth]{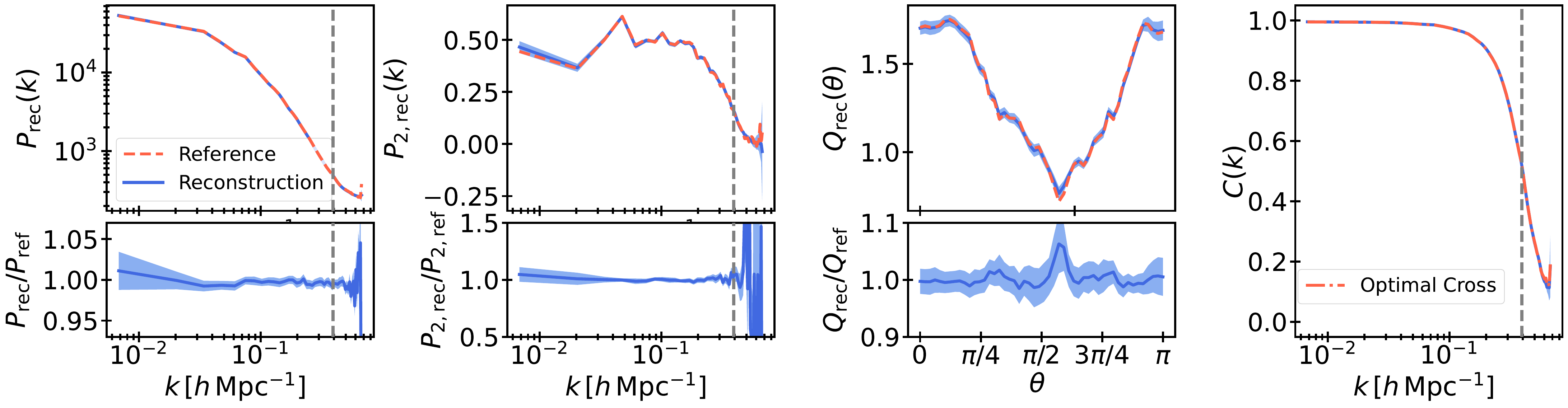}
    \caption{\tthree:  Final conditions. Analogous to Fig.~\ref{fig:ntr_stats}.}\label{fig:ntr_stats5HIC}
\end{figure*}

The additional numerical case \tthree in which the bias parameters are assumed to be known, allows us to implement the most complex bias model as in \citet{Coloma_2024}. Specifically, we define distinct bias prescriptions across the hierarchical cosmic-web, using a classification based on both the tidal field tensor and the Hessian of the density field. This results in 16 regions, each characterised by its own power-law, exponential threshold, and stochastic bias parameters. As shown in Figs.~\ref{fig:maps8} \ref{fig:din_stats5HIC}, and \ref{fig:ntr_stats5HIC}, the reconstruction maintains the same level of precision under this more sophisticated setting, as in \tone and \ttwo{,  assuming, however, the bias parameters to be known}.


\end{appendix}

\end{document}